\documentclass[structabstract]{aa}

\usepackage{amsmath}
\usepackage{txfonts}
\usepackage{booktabs}
\usepackage{array}
\usepackage{graphicx}
\usepackage{natbib,twoopt}
\usepackage{upgreek}
\usepackage{color}
\usepackage[breaklinks=true]{hyperref} %% to avoid \citeads line fills
\bibpunct{(}{)}{;}{a}{}{,} %% natbib format like A&A and ApJ
\newcommandtwoopt{\citeads}[3][][]{\href{http://adsabs.harvard.edu/abs/#3}%
{\citealp[#1][#2]{#3}}}
\newcommandtwoopt{\citepads}[3][][]{\href{http://adsabs.harvard.edu/abs/#3}%
{\citep[#1][#2]{#3}}}
\newcommandtwoopt{\citetads}[3][][]{\href{http://adsabs.harvard.edu/abs/#3}%
{\citet[#1][#2]{#3}}}
\newcommandtwoopt{\citeyearads}[3][][]%
{\href{http://adsabs.harvard.edu/abs/#3}{\citeyear[#1][#2]{#3}}} 

\begin{document}

\title{\object{M82} - A radio continuum and polarisation study}
\subtitle{II. Polarisation and rotation measures}
\author{B. Adebahr\inst{1,2,3}\and M. Krause\inst{2} \and U. Klein\inst{4} \and G. Heald\inst{3,5,6} \and R.-J. Dettmar\inst{1}}
\institute{Astronomisches Institut der Ruhr-Universit\"at Bochum (AIRUB), Universit\"atsstrasse 150, 44801 Bochum
\and Max-Planck-Institut f\"ur Radioastronomie (MPIfR), Auf dem H\"ugel 69, 53121 Bonn
\and ASTRON, PO Box 2, 7990 AA Dwingeloo, The Netherlands
\and Argelander-Institut f\"ur Astronomie (AIfA), Universit\"at Bonn, Auf dem H\"ugel 71, 53121 Bonn, Germany
\and CSIRO Astronomy and Space Science, 26 Dick Perry Avenue, Kensington WA 6151, Australia
\and Kapteyn Astronomical Institute, University of Groningen, PO Box 800, 9700 AV, Groningen, The Netherlands}

\date{-}
\abstract {The composition and morphology of the interstellar medium in starburst galaxies has been well investigated, but the magnetic field properties are still uncertain. The nearby starburst galaxy M82 provides a unique opportunity to investigate the mechanisms leading to the amplification and reduction of turbulent and regular magnetic fields.} {An investigation of the magnetic field properties in M82 will give insight into mechanisms to generate and maintain a magnetic field as well as depolarisation mechanisms. Possible scenarios of the contribution of the magnetic field to the star-formation rate are evaluated.} {Archival data from the Very Large Array (VLA) were combined and re-reduced to cover the wavelength regime at $\uplambda3$\,cm and $\uplambda6$\,cm. Complementary data from the Westerbork Synthesis Radio Telescope (WSRT) at $\uplambda18$\,cm and $\uplambda22$\,cm were reduced and analysed using the RM-Synthesis technique.} {All observations revealed polarised emission in the inner part of the galaxy, while extended polarised emission up to a distance of 2\,kpc from the disk was only detected at $\uplambda18$\,cm and $\uplambda22$\,cm. The observations hint at a magnetised bar in the inner part of the galaxy. We calculate the mass inflow rate due to magnetic stress of the bar to 7.1\,M$_\odot$yr$^{-1}$, which can be a significant contribution to the star-formation rate (SFR) of M82 of $\sim13$\,M$_\odot$yr$^{-1}$. The halo shows polarised emission, which might be the remnant of a regular disk field. Indications for a helical field in the inner part of the outflow cone are provided. The coherence length of the magnetic field in the centre could be estimated to 50\,pc, which is similar to the size of giant molecular clouds. Using polarisation spectra more evidence for a close coupling of the ionised gas and the magnetic field as well as a two-phase magnetic field topology were found. Electron densities in the halo ($\langle n_e \rangle\approx0.009$\,cm$^{-3}$) are similar to the ones found in the Milky Way.} {The magnetic field morphology is similar to the one in other nearby starburst galaxies (NGC1569, NGC253) with possible large-scale magnetic loops in the halo and a helical magnetic field inside the outflow cones. The special combination of a magnetic bar and a circumnuclear ring are able to significantly raise the star-formation rate in this galaxy by magnetic braking, but cannot be the cause for all starbursts.}
\maketitle

\section{Introduction}

Magnetic fields are an important ingredient to explain the physics of galaxies and their evolution over the entire age of the Universe. The magnetic field generation and morphology in starburst galaxies however remains unclear, even though this type of galaxy might be one of the most important contributors to the magnetisation of the Universe during its early stages \citep{1999ApJ...511...56K,2006MNRAS.370..319B} due to its high star-formation rate and therefore strong outflows. The key to understanding these outflow dynamics and their composition is observations of the different phases of the interstellar medium and the consequent knowledge of its kinematics, densities, and morphologies.

Analysis of the composition and interaction of the various phases of the interstellar medium in nearby starbursting galaxies by, for example, \citet{2004ApJS..151..193S} and \citet{2015ApJ...814...83L} could constrain the distribution and physical parameters on different scales, but the role of magnetic fields, active or passive, remains unknown. This is mostly due to insufficient spatial resolution and sensitivity to disentangle the complicated structures and physics. Acquiring knowledge of the evolution of these parameters during an interaction and the following starbursting process, and what ingredients are needed to amplify star formation and magnetic fields, are inevitably important for our understanding of the magnetic-field formation in the early Universe. \citet{2011MNRAS.tmp.2083G} and \citet{2011MNRAS.415.3189K} already showed that minor and major mergers contribute to the star formation as well as to the magnetic field amplification significantly.

The usual explanation for the observed regular magnetic field morphology in galaxies is the $\alpha\omega$-dynamo in combination with a galactic wind driven by star formation and subsequent supernova explosions \citep{1996ARA&A..34..155B,2009RMxAC..36...25K}. Such a dynamo needs a buildup time of at least 6\,Gyrs for dwarf galaxies and even longer for normal spirals to generate regular magnetic field patterns \citep{2009A&A...494...21A}. Recently, \citet{2014A&A...566A..40M} showed that a dynamo can exist even during periods of violent starburst activity and can be enhanced by tidal interactions. This still does not explain the detection of regular magnetic fields in the halos of galaxies at high redshifts \citep{2008Natur.454..302B}. However, \citet{2010A&A...523A..72D} presented a scenario where the first dwarf galaxies were able to generate those field strengths in the early Universe, referring to this phenomenon as a cosmic dynamo.

Therefore questions arise from where and how these magnetic fields are generated: Is a primordial seed field in combination with a dynamo mechanism needed to generate the first magnetic fields and have they lead to the magnetic field morphology of nearby galaxies at the current epoch \citep{1988PhRvD..37.2743T}? Or could the first supernova explosions have driven this process and magnetised intergalactic space \citep{2009A&A...494...21A}? With the well studied primordial starburst galaxy M82 being very close-by, we have the unique opportunity to examine in much more detail the morphology of magnetic fields in an environment similar to the one in the early Universe.

M82 is a nearby starbursting galaxy at a distance of $\sim3.5$\,Mpc \citep{2009AJ....138..332J}. Observations at several different wavelengths have revealed the distribution and kinematics of the different ISM gas phases (\ion{H}{i}: \citet{1993ApJ...411L..17Y}, CO: \citet{2002ApJ...580L..21W}, H$_2$: \citet{2009ApJ...700L.149V}, dust: \citet{2002PASJ...54..891O,2009AJ....137..517L}) as well as the extent of the bipolar outflow originating from the starbursting centre (SFR$\approx13$\,M$_\odot$yr$^{-1}$ \citep{2003ApJ...599..193F}) of the galaxy (X-rays: \citet{2009ApJ...697.2030S}, H$\upalpha$: \citet{1999ApJ...510..197D, 2002PASJ...54..891O}, radio continuum: \citet{2013A&A...555A..23A, 1994A&A...282..724R, 1992A&A...256...10R}). This starburst could have been triggered by an interaction with the grand-design spiral galaxy M81 \citep{2008AJ....135.1983C}. While the first interaction between these two galaxies happened around 100\,Myrs ago, the last starburst periods of M82 are known to have occured 5-10\,Myrs ago \citep{2003ApJ...599..193F}.

While the morphology of this galaxy was originally claimed as irregular \citep{1979AJ.....84..472S}, studies by \citet{1991ApJ...369..135T} and \citet{2002A&A...383...56G} revealed a stellar bar with a length of 1\,kpc in the centre. \citet{2005ApJ...628L..33M} could even find spiral arms connected to the tips of this stellar bar and claim that M82 has lost a significant part of its mass due to its encounter with M81 hinting at a former small spiral morphology.

The aim of this paper is to examine the magnetic field morphology of the starburst galaxy M82 and draw conclusions on its possible origins as well as discussing their impact on the evolution of starburst galaxies. For this publication we used radio continuum polarisation observations at four different wavelengths between $\uplambda3$\,cm and $\uplambda22$\,cm. The derived polarised intensities and rotation measures (RMs) allow us to estimate regular and turbulent magnetic field strengths and derive thermal electron densities inside the polarised structures. This enables us to characterise not only the average density of the medium, but also to draw conclusions about its filling factor, morphology, and connection to magnetic fields.

In Section \ref{text_datareduction} we describe the data reduction techniques and the differential technique we used to analyse our RM-cubes in order to reach the needed dynamic range. Sections \ref{text_polem3_6} and \ref{Sect:18/22} describe the morphology of the detected polarised emission and the RMs. We draw conclusions on the overall magnetic field morphology of M82 in Section \ref{text_magfield_struc}. Using an equipartition assumption, we calculate the turbulent and regular magnetic field strengths in the detected polarised structures in Section \ref{text_magfieldstr}. In Section \ref{text_depol} we use the detected degrees of polarisation in the galaxy itself as well as in a background source to identify the dominant depolarisation mechanism and depolarising morphology. We discuss our results in Section \ref{text_discussion} and summarise them in Section \ref{text_summary}.

\section{Observations and data reduction}
\label{text_datareduction}

The observational parameters of the data with full polarisation information were already described by \citet{2013A&A...555A..23A}, therefore we only describe the most relevant observational details and data-reduction steps here.

\subsection{VLA data reduction}

The VLA observations are two archival data sets at $\uplambda3$\,cm and $\uplambda6$\,cm already published by \citet{1991ApJ...369..320S} and \citet{1992A&A...256...10R,1994A&A...282..724R}. However, the polarisation information was only analysed for one of these data sets.

The $\uplambda3$\,cm observations consist of two 50\,MHz sub-bands at 8465\,MHz and 8415\,MHz with a total integration time of 160 minutes and the $\uplambda6$\,cm observations of two 50\,MHz sub-bands at 4885\,MHz and 4835\,MHz with a total integration time of 385 minutes. For both observations only the data from the compact VLA-D array configuration were used to improve the sensitivity to large structures.

For both data sets, the data reduction and imaging were done entirely in CASA (Common Astronomy Software Applications) \citep{2007ASPC..376..127M} using the time-dependant gain, polarisation leakage and polarisation-angle corrections from the calibrator sources 3C138, 3C286, 3C48, 1044+719, and 0836+710 after careful editing for radio frequency interference (RFI). Self-calibration was done in full polarisation for each sub-band individually using solutions derived from clean components with masks from the total power image. For the final imaging, the observations were combined and one image per sub-band was produced. Images at the same observing band were then combined with an inverse variance weighting and corrected for the primary beam response yielding one image each in Stokes Q and U at $\uplambda3$\,cm and $\uplambda6$\,cm.

From these images polarised intensity and polarisation angle images were produced for each observing band with MIRIAD (Multichannel Image Reconstruction Image Analysis and Display) \citep{1995ASPC...77..433S}. The polarisation bias was subtracted using the noise in the individual Stokes Q and U images. Polarisation images were blanked below a signal-to-noise level of 2 and polarisation angle images at pixels where the error reached values of more than 10$^\circ$. To produce the RM image, the $\uplambda3$\,cm Q and U images were first convolved to the $\uplambda6$\,cm beam. Then an RM was determined for each individual pixel.

The final polarisation maps may suffer from depolarisation inside the observing band, if the polarisation vectors rotate very fast due to high RMs. For a complete depolarisation, this occurs if the vector rotates by 90$^\circ$ inside the observing band, which would occur for RMs of $10^5$\,rad\,m$^{-2}$ at $\uplambda3\,$cm and of $2\cdot10^4$\,rad\,m$^{-2}$ at $\uplambda6\,$cm, respectively. RMs of this magnitude are only known to occur in the strongest polarised active galactic nuclei (AGN) and the effect is therefore negligible for this study. Another effect is the $n\pi$-ambiguity, which occurs if the vector rotates by 180$^\circ$  $n$ times between two observing bands resulting in an underestimate of the RMs in the final maps. This would be the case for RMs of $\sim n\cdot1200$\,rad\,m$^{-2}$. Since RMs of such an order have never been observed in any nearby galaxy, values of $|n|>1$ would be very unexpected.

\subsection{WSRT data reduction}

The WSRT observations are unpublished data from the WSRT-SINGS survey, so the observational setup will not be repeated here (see \citet{2007A&A...461..455B}). The calibration strategy was improved using the following steps. First the system temperatures were applied and the datasets concatenated inside the AIPS software package. Special care was taken when handling the polarisation products since the WSRT uses linear feeds at L-band while AIPS assumes circular ones. The data were then converted to CASA Measurement Set (MS) format for flagging. This was done with a specially tailored strategy for this observation using the semi-automated flagging routine RFIconsole \citep{2010MNRAS.405..155O}. After this the calibrator gains as well as the polarisation leakage and angle corrections were applied on an individual channel basis using the calibrator sources CTD93 and 3C138. This minimises the impact of the standing wave inside the WSRT dishes, which causes a resonance with a period of 17\,MHz \citep{2005A&A...441..931D,2008A&A...489...69B}.

For self-calibration, the dataset was exported to MIRIAD. Because of the high RMs on the order of $|RM|\approx600$\,rad\,m$^{-2}$ already detected in M82 by \citet{1994A&A...282..724R}, which would cause bandpass depolarisation, each channel was imaged and self-calibrated in all four Stokes parameters individually. This was of utmost importance for the polarisation analysis since the WSRT-SINGS frequency setup spans two individual bands with a bandwidth of 160\,MHz each, over a full bandwidth of $\approx450$\,MHz. Any imaging of the full bandwidth in one imaging step would have even depolarised structures with small RM-values.

The cleaning of the individual images was done with a mask from the total power emission to limit the algorithm to real emission only. For each self-calibration cycle the clean iterations were increased and the solution interval decreased until a 3$\sigma$-noise level and the integration time of one minute was reached. Final images for each Stokes Q and U parameter and for each individual channel were produced by cleaning down to a 2$\sigma$-noise level and applying the primary beam correction. Each individual image was then checked for its quality and any obviously bad images were excluded from further data reduction and analysis steps. This yielded only 396 (187 for the $\uplambda18$\,cm and 209 for the $\uplambda22$\,cm observations) out of 1024 final images for each polarisation product because of the large amount of RFI and strong solar activity during the observation.

\subsubsection{Rotation-measure-Synthesis}

In contrast to the classical approach that was applied to the archival VLA data, the WSRT with its broadband multichannel backend provides extended capabilities for the analysis of polarisation data. The small single channel bandwidth of 312.5\,kHz and the large number of available channels combined with the broad bandwidth of 160\,MHz allows the use of the RM-Synthesis technique, which was analytically described by \citet{1966MNRAS.133...67B} for single lines of sight and extended to diffuse emission by \citet{2005A&A...441.1217B}.

\begin{figure}
        \resizebox{\hsize}{!}{\includegraphics{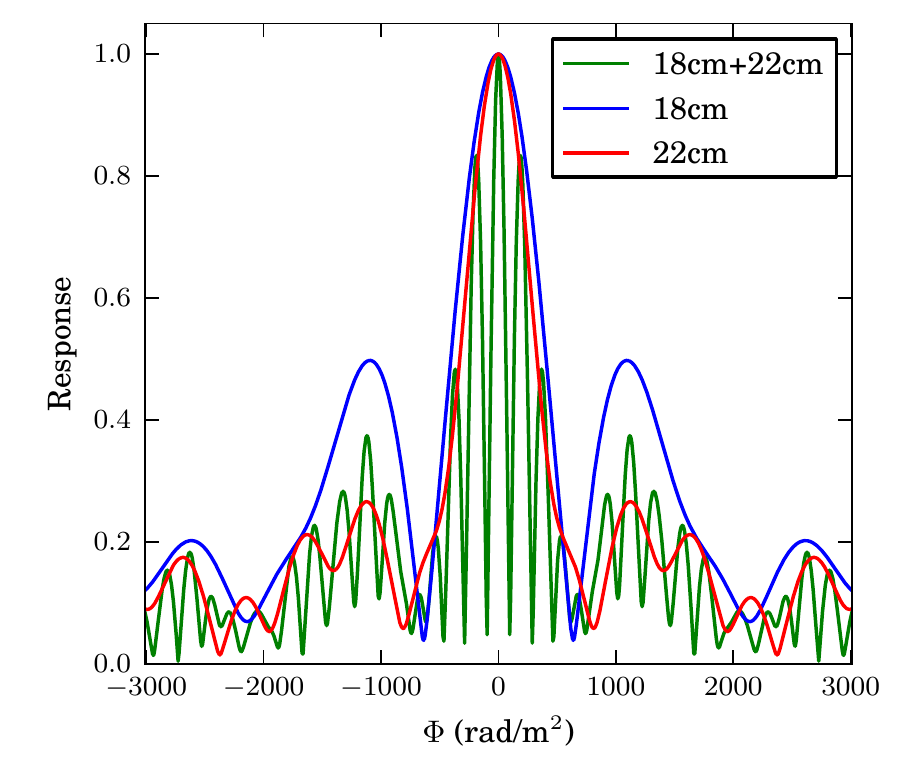}}
        \caption{Amplitude response of the RM spread
function (RMSF) for the three different data combinations of the RM-Synthesis. All amplitudes are normalised to unity.}
        \label{plot_rmsf}
\end{figure}

This technique was conducted on the final $\uplambda18$\,cm and $\uplambda22$\,cm observing bands individually and on a combination of both using the code from \citet{2005A&A...441..931D}. All images were first convolved to the size of the beam of the lowest frequencies (15\farcs7$\times$14\farcs2) using the MIRIAD task CONVOL. In addition we produced a second set of images convolved to a beam approximately three times larger (43\farcs1$\times$39\farcs6) in size to increase the sensitivity to diffuse polarised emission.

Each RM cube was sampled between $-3000$\,rad\,m$^{-2}$ and $3000$\,rad\,m$^{-2}$ with a step size of 3\,rad\,m$^{-2}$ to search for high RMs. The RMSFs are plotted in Fig. \ref{plot_rmsf}. Due to the frequency gap between the $\uplambda18$\,cm and $\uplambda22$\,cm bands, the sidelobes in the final RM-cubes using both wavelengths ranges together reach values of $\sim80\%$ of the main peak. This can lead to confusion between sidelobes and spectra with multiple peaks. Therefore the cubes were cleaned using the RM-clean algorithm \citep{2009A&A...503..409H} down to the 3$\sigma$-noise level of the individual Stokes Q and U cubes after running RM-Synthesis.

The calculated parameters for the RM-Synthesis are listed in Table \ref{table_obs}. The FWHM of the RM spread function of the observation is always larger than its scale. This makes polarised structures with a wider extent in Faraday Space than the largest observable scale look like several peaks accompanied by missing emission in between them \citep{2009A&A...503..409H}. This problem is similar to the missing spacing problem for interferometric imaging of objects with large angular extents.

\begin{table}
        \caption{Parameters for the RM-Synthesis. RM$_{\text{max}}$ is the largest detectable RM, FWHM is the full width at half maximum of the RMSF, and Scale is the widest detectable structure in RM.}
        \label{table_obs}
        \centering
        \newcolumntype{1}{>{\centering\arraybackslash} m{1.4cm} }
        \newcolumntype{2}{>{\centering\arraybackslash} m{1.3cm} }
        \newcolumntype{3}{>{\centering\arraybackslash} m{1.4cm} }
        \newcolumntype{4}{>{\centering\arraybackslash} m{1.4cm} }
        \begin{tabular}{@{} 1 2 3 4 @{}}
                \toprule
                \toprule
                Wavelength (cm) & RM$_{\text{max}}$ (rad/m$^2$) & FWHM (rad/m$^2$) & Scale (rad/m$^2$) \\
                \midrule
                18 & $1.5\cdot 10^5$ & 854 & 106 \\
                22 & $0.8\cdot 10^5$ & 542 & 72 \\
                18+22 & $1.1\cdot 10^5$ & 167 & 106 \\
                \bottomrule
        \end{tabular}
\end{table}

\subsection{Instrumental effects and polarisation bias}
\label{text_instpol_polbias}

In the case of a strong source like M82, the final polarisation images can be influenced by instrumental polarisation artefacts resulting in polarised emission at the locations of strong sources in total emission. For our VLA data this was discussed in detail by \citet{1994A&A...282..724R}, who came to the conclusion that no instrumental polarisation is influencing the data. To quantify their statement, the polarised flux of the unpolarised calibrator can be measured and compared to its flux in total power. The ratio between the two values can then serve as a limit for on-axis targets for the fractional polarisation up to which a polarised signal is significantly of astronomical origin. 

No signal in Stokes Q or U could be detected for the unpolarised calibrators of the VLA $\uplambda$3\,cm or $\uplambda$6\,cm observations. Using the 3$\sigma$ rms (where $\sigma$ is the rms level) of the calibrator maps as an upper limit leads to reliable polarisation percentages of 0.03\,\%. By comparing the highest values in the total power maps of M82 with those limits resulted in an upper limit for the instrumental contribution to the fluxes of 35\,$\upmu$Jy at $\uplambda$3\,cm and 83\,$\upmu$Jy at $\uplambda$6\,cm for the polarised intensity maps. The polarisation bias for these observations was then subtracted using the routine implemented in the MIRIAD task IMPOL, while the RMs and intrinsic magnetic field angles were calculated using the task IMRM.

Typical instrumental polarisation contributions for the WSRT are below a 3$\sigma$ level of 0.5\,\% for sources in the pointing centre if the channel-based application of the calibrator gains was used \citep{2011A&A...526A...9B}. To verify this further, RM-Synthesis was also conducted on the unpolarised calibrators used for the observations, which are used to minimise the leakage of the total power emission into the polarised signal. The polarised signal in the final RM cubes reached a $1\sigma$-level of 0.05\,\%, which still leads to instrumental polarisations of ~700\,$\upmu$Jy for the centre of the galaxy and even ~100\,$\upmu$Jy for the disk region, which is larger than our noise in the individual RM cubes of ~15\,$\upmu$Jy.

This instrumental polarisation is visible in the shape of a butterfly centred at the position of the strongest total power emission. This is typical for an instrumental artefact and was described for the WSRT by \citet{2005A&A...441..931D}. Since instrumental polarisation is retrieved as a constant flux over frequency in Stokes Q and U, it appears as a single peak at an RM=0\,rad\,m$^{-2}$, that can be mitigated. Instrumental polarisation also causes sidelobes in the RM spectra and can be cleaned with the RM-Clean algorithm, but due to the high sidelobe levels we could not remove all the sidelobe flux from the spectra. This might be caused by the dependence of the shape of the RMSF on the spectral index of the polarised emission over the observed frequency \citep{2005A&A...441.1217B}. The result are residuals of the RMSF in the resulting RM spectra which are symmetric around RM$_0$.

To mitigate instrumental polarisation in our $\uplambda18$\,cm and $\uplambda22$\,cm data sets we applied the differential approach introduced by \citet{2013A&A...559A..27G}, which uses the instrumental polarisation at RM=0\,rad\,m$^{-2}$ to produce a symmetric pattern at positive and negative RMs. A cleaned RM cube mirrored along the RM axis around RM=0\,rad\,m$^{-2}$ was subtracted from the original cleaned RM cube. This removes all instrumental polarisation at RM=0\,rad\,m$^{-2}$. An obvious disadvantage is that real signals close to 0\,rad\,m$^{-2}$, or symmetric features in Faraday depth spectra (like those observed by \citet{2009A&A...503..409H}) will also be strongly reduced in strength by this method. When we apply the differential technique by \citet{2013A&A...559A..27G} to our data, the wide RMSF means that we cannot detect polarised emission out to several hundred rad\,m$^{-2}$ in the core of M82. Therefore we can only investigate the outskirts and the halo of this galaxy. Instrumental polarisation for the WSRT is strongly frequency- and direction dependent and can reach ratios of 5\%-10\% \citep{2008A&A...479..903P} at a distance of $\approx30\arcmin$ from the pointing centre. Since the maximum distance of our detected structures is $\leq3\arcmin$ we do not expect any instrumental polarisation contributing to the halo emission.

Extended diffuse emission can be affected by missing spacings (the absence of short baselines of the observing telescope array resulting in lower flux measurements). We have successfully shown that this is not the case for our total flux measurements in \citet{2013A&A...555A..23A}, which were unaffected up to extents of $2\arcmin$ at $\uplambda3$\,cm/$\uplambda6$\,cm and $6\arcmin$ at $\uplambda22$\,cm. Our polarised emission shows smaller extents in all cases and therefore we conclude that missing spacings are not affecting our measurements.

To correct for polarisation bias from the RM cubes, we calculated the mean value of the polarised intensity along each line of sight in RA-DEC-space excluding all obviously visible emission. A parabola was then fitted to the distance of each pixel from the pointing centre and its value. Using the fitted value from this parabola in all directions from the pointing centre produces a paraboloid resembling the polarisation bias. This was then subtracted from the polarised intensity cube.

\section{Polarised emission at $\uplambda$3\,cm and $\uplambda$6\,cm}
\label{text_polem3_6}

\subsection{Morphology}

\begin{figure}
        \resizebox{\hsize}{!}{\includegraphics{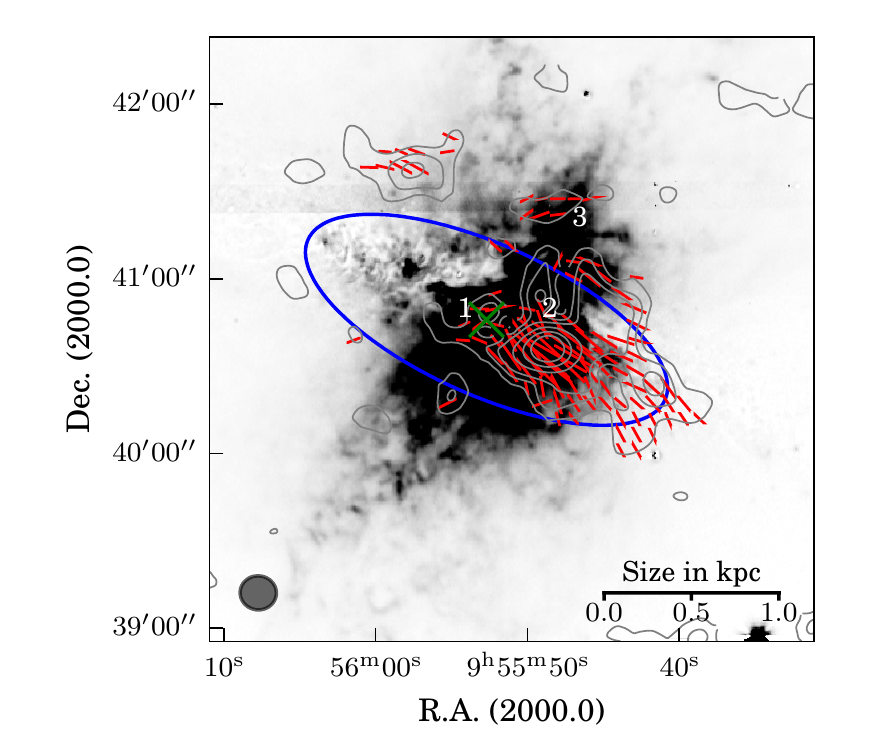}}
        \caption{Polarised flux density contours at $\uplambda$3\,cm smoothed to the $\uplambda$6\,cm resolution overlaid on a H$\upalpha$ image from the Wisconsin Indiana Yale National (WIYN) Observatory. Contour levels are at 100, 150, 200, 300, 400, and 500\,$\upmu$Jy beam$^{-1}$. In the lower left we show the size of the synthesised beam (12\farcs5$\times$11\farcs7). The red vectors indicate the measured polarised flux density ($1\arcsec$ = 40\,$\upmu$Jy beam$^{-1}$) and the intrinsic magnetic field direction of the emission. The green cross marks the optical centre of M82 and the blue ellipse the extent of the central optical disk.}
        \label{image_MF_3cm}
\end{figure}

\begin{figure}
        \resizebox{\hsize}{!}{\includegraphics{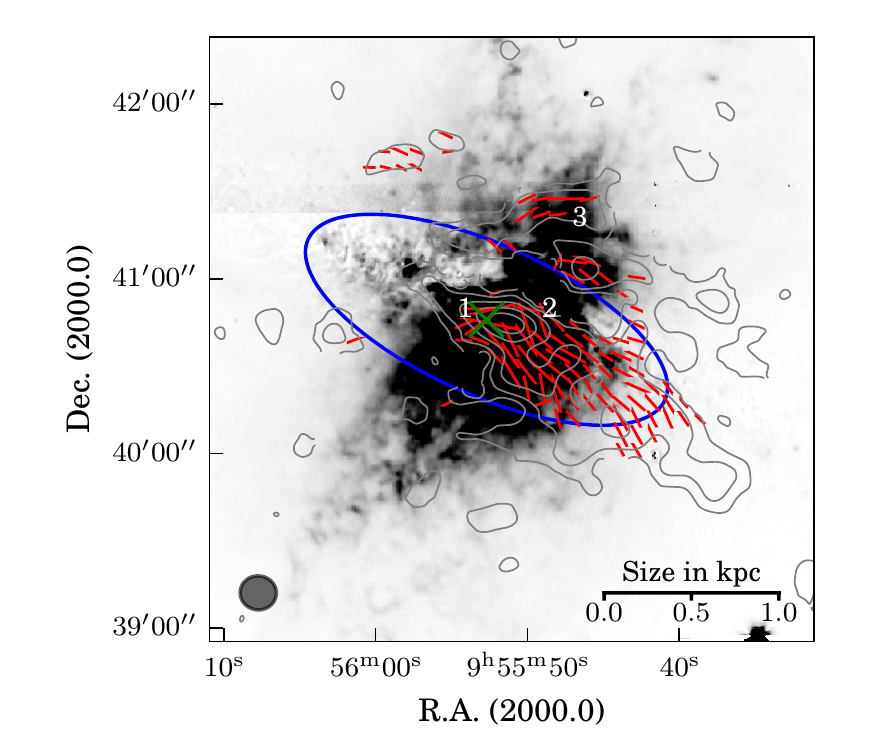}}
        \caption{Polarised flux density contours at $\uplambda$6\,cm overlaid on a H$\upalpha$ image from the WIYN Observatory. Contour levels are at 80, 120, 200, 300, 400\,$\upmu$Jy beam$^{-1}$. In the lower left we show the size of the synthesised beam (12\farcs5$\times$11\farcs7). The red vectors indicate the measured polarised flux density ($1\arcsec$ = 40\,$\upmu$Jy beam$^{-1}$) and the intrinsic magnetic field direction of the emission. The green cross marks the optical centre of M82 and the blue ellipse the extent of the central optical disk.}
        \label{image_MF_6cm}
\end{figure}

Nearly the entire inner disk of M82 emits polarised emission at $\uplambda$3\,cm (Fig. \ref{image_MF_3cm}) and $\uplambda$6\,cm (Fig. \ref{image_MF_6cm}), with some diffuse emission extending towards the northern halo and western disk. The overall morphology of this diffuse emission shows no correlation with the H$\upalpha$ emission. While the strongest emission in H$\upalpha$ is visible towards the south, the polarised emission extends preferentially towards the north. At both radio wavelengths an asymmetry is visible in the disk. While the emission extends 1.3\,kpc ($\uplambda$3\,cm) and 1.9\,kpc ($\uplambda$6\,cm) towards the west (Region 2), nearly no emission shows up towards the east. An offset of the strongest emission is visible in the radio as well as the H$\upalpha$ maps $\sim0.2$\,kpc towards the east in comparison to the optical centre of the galaxy. The maps at both wavelengths show no polarised signal inside the outflow cone, about 0.4\,kpc from the centre, but further into the halo the polarised emission shows up again at a distance of 0.6\,kpc (Region 3). The non-detection of polarised emission between these regions is most likely caused by beam depolarisation due to the superposition of two components of the magnetic field in different regions within the same resolution element; one parallel to the major axis in the bar and the other perpendicular to the disk following the main outflow direction. A remarkable region of diffuse polarised emission is detected towards the northwest and above and below the western disk in a plume-like shape. The centre of the galaxy also shows polarised emission (Region 1). The instrumental polarised flux density in this area is at least a factor of 3 lower than the detected flux densities (see Sect. \ref{text_instpol_polbias}), so that the detected emission is reliable.

\subsection{Rotation measures}

\begin{figure}
        \resizebox{\hsize}{!}{\includegraphics{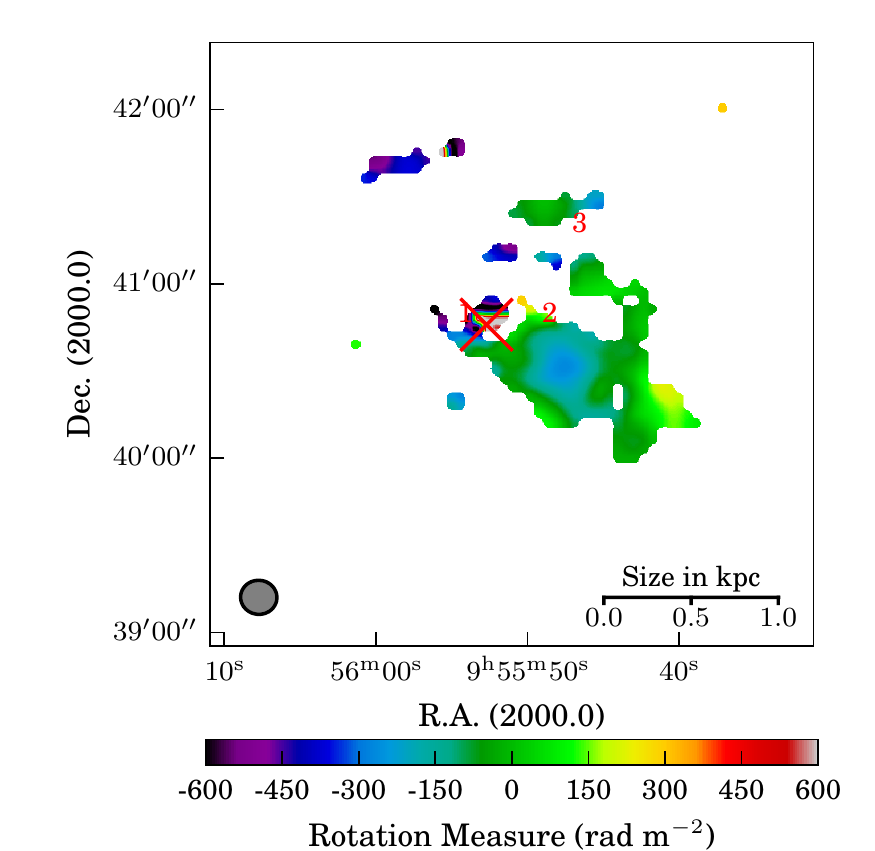}}
        \caption{RMs between $\uplambda$3\,cm and $\uplambda$6\,cm above a $3\sigma$ detection limit for each of the two polarised flux density maps used. In the lower left we show the size of the synthesised beam (12\farcs5$\times$11\farcs7). The galactic foreground RM of -20\,rad\,m$^{-2}$ \citep{2012A&A...542A..93O} was corrected for. The red cross marks the optical centre of the galaxy.}
        \label{image_RM_3cm_6cm}
\end{figure}

The calculated RMs between $\uplambda$3\,cm and $\uplambda$6\,cm were corrected for the RM of the Galactic foreground using the value from \citet{2012A&A...542A..93O} of $-19.65\pm5.95$\,rad\,m$^{-2}$. The resulting map is shown in Fig. \ref{image_RM_3cm_6cm}. Errors for this map are on the order of $\pm30$\,rad\,m$^{-2}$ for the centre and of $\pm65$\,rad\,m$^{-2}$ for the diffuse emission close to the detection limit.

The detected RMs show a smooth cuspy distribution across the western part of the galaxy with values ranging from $-100$\,rad\,m$^{-2}$ to $-300$\,rad\,m$^{-2}$. This smooth morphology shows that the $n\pi$-ambiguity is not important in our observations, since $|n|>1$ would be visible as sudden jumps in RM over distances the size of the synthesised beam. Values of $n>1$ would cause the RMs to change signs and vary by large values on scales of the beam.

RMs between $-600$\,rad\,m$^{-2}$ and 600\,rad\,m$^{-2}$ are detected in the centre. Region 3 in the northern halo of M82 shows RMs between $-100$\,rad\,m$^{-2}$ and $+50$\,rad\,m$^{-2}$; the intrinsic magnetic field vectors in this region bend across the emission region. Higher absolute RMs are visible towards the centre of the H$\upalpha$ outflow cone.

Despite different reduction and analysis procedures, our results are very similar to that in \citet{1994A&A...282..724R} \footnote{The orientation of the images in \citet{1994A&A...282..724R} was flipped. This was later corrected in the Erratum \citep{1995A&A...293..287R}}.

\section{Polarised emission at $\uplambda$18\,cm and $\uplambda$22\,cm}
\label{Sect:18/22}

The RM cube of the combined $\uplambda$18\,cm and $\uplambda$22\,cm bands was investigated for peaks along the Faraday depth axis. We could not find any reliable emission with $|RM|>600$\,rad\,m$^{-2}$, therefore we attribute any peaks in this range as noise or remnants of the subtraction of the instrumental polarisation. The position of a peak in the RM cube directly gives the RM of the observed polarised component while the amplitude of the peak corresponds to the polarised intensity. To calculate the intrinsic magnetic field orientation we used
\begin{equation}
        \Theta_0 = \Theta - RM \cdot \lambda_0^2
,\end{equation}
with
\begin{equation}
        \Theta = \frac{1}{2}\arctan \left(\frac{U}{Q}\right) + \frac{\pi}{2},
\end{equation}
where $\Theta_0$ is the intrinsic and $\Theta$ the observed magnetic field angle counted from north to east in rad, $RM$  the RM at the position of the peak, $\lambda_0^2$ the weighted mean wavelength squared of the frequency band, and $U$ and $Q$ are the corresponding fluxes in the Stokes U- and Q cubes at the position of the peak in the polarised intensity cube. The resulting RM- and $\Theta_0$-maps were also corrected for Galactic Foreground rotation. This value was also cross checked using our RM-Synthesis maps and two polarised point-sources close to the galaxy, for which the RMs are $-25.44\pm0.16$\,rad\,m$^{-2}$ and $-27.23\pm1.07$\,rad\,m$^{-2}$; those are slightly more negative, but might be contaminated by fields inside the M81-M82-group itself or outskirts of the halo of M82, therefore we assumed a value of $-19.65$\,rad\,m$^{-2}$ throughout this paper. We calculated $RM$ and $\Theta_0$ for each individual line of sight for any pixel with a signal-to-noise ratio of more than five times the noise of the individual Stokes U- and Q cubes.

\subsection{Morphology}

\begin{figure}
        \resizebox{\hsize}{!}{\includegraphics{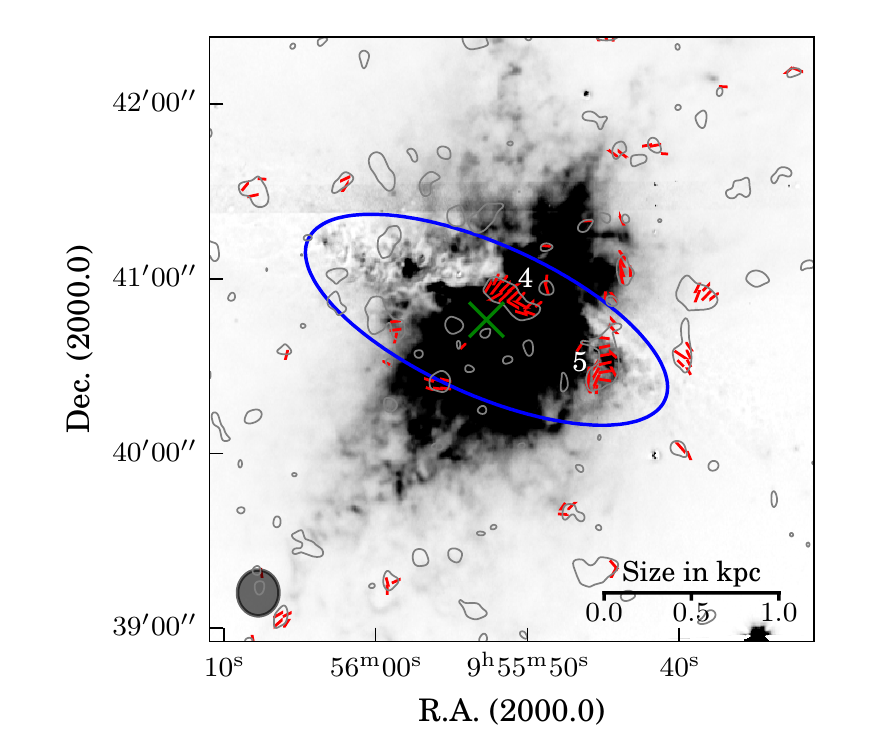}}
        \caption{Polarised flux density contours from the RM cube of the combined $\uplambda$18\,cm and $\uplambda$22\,cm data using the differential technique overlaid on a H$\upalpha$ image from the WIYN Observatory. The contour level is at 65\,$\upmu$Jy beam$^{-1}$, which corresponds to a 5$\sigma$ noise level in the individual Stokes Q and U cubes. In the lower left we show the size of the synthesised beam (15\farcs7$\times$14\farcs2). The red vectors indicate the measured polarised flux density ($1\arcsec$ = 40\,$\upmu$Jy beam$^{-1}$) and the intrinsic magnetic field direction of the emission. The green cross marks the optical centre of M82 and the blue ellipse the extent of the central optical disk.}
        \label{image_18cm_22cm_sm}
\end{figure}

\begin{figure}
        \resizebox{\hsize}{!}{\includegraphics{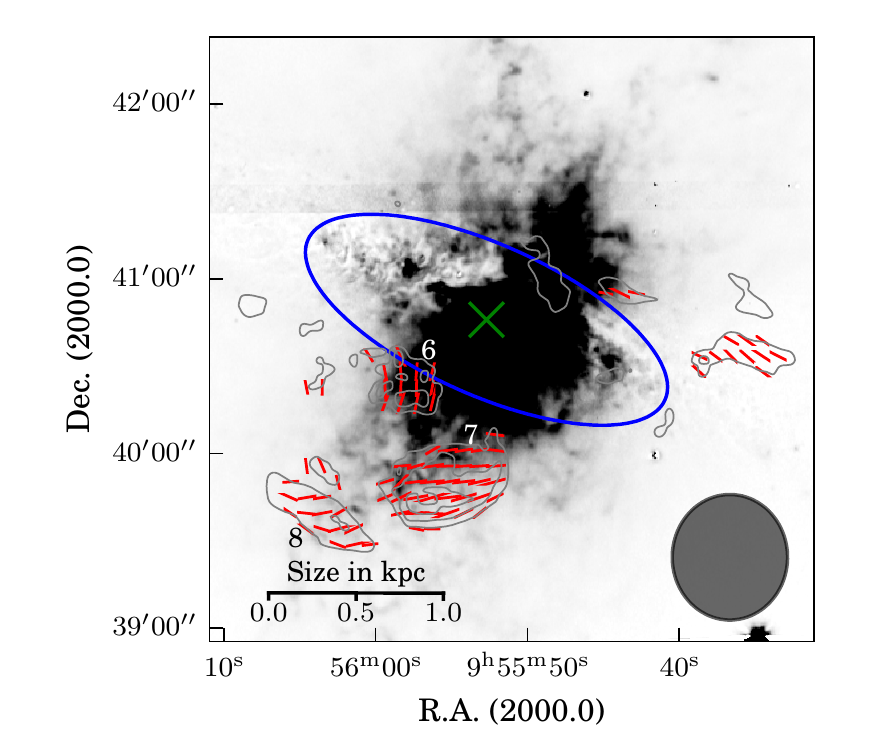}}
        \caption{Polarised flux density contours from the smoothed RM cube of the combined $\uplambda$18\,cm and $\uplambda$22\,cm data using the differential technique overlaid on a H$\upalpha$ image from the WIYN Observatory. The contour levels are at 120, 160, and 200\,$\upmu$Jy beam$^{-1}$, which corresponds to a 5$\sigma$ noise level in the individual Stokes U- and Q cubes. In the lower right we show the size of the synthesised beam (43\farcs1$\times$39\farcs6). The red vectors indicate the measured polarised flux density ($1\arcsec$ = 40\,$\upmu$Jy beam$^{-1}$) and the intrinsic magnetic field direction of the emission. The green cross marks the optical centre of M82 and the blue ellipse the extent of the central optical disk.}
        \label{image_MF_18cm_22cm_lg_sublr}
\end{figure}

\begin{figure}
        \resizebox{\hsize}{!}{\includegraphics{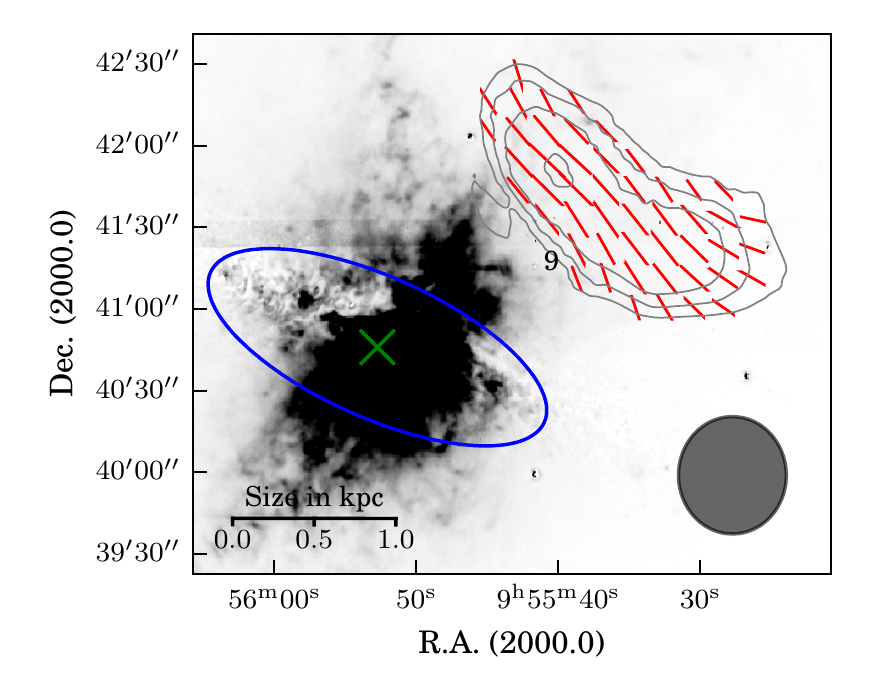}}
        \caption{Polarised flux density contours from the smoothed RM cube of the combined $\uplambda$18\,cm and $\uplambda$22\,cm data overlaid on a H$\upalpha$ image from the WIYN Observatory. The contour levels are at 200, 250, 300, and 400\,$\upmu$Jy beam$^{-1}$, which corresponds to a 5$\sigma$ noise level in the individual Stokes U- and Q cubes. In the lower left we show the size of the synthesised beam (43\farcs1$\times$39\farcs6). The red vectors indicate the measured polarised flux density ($1\arcsec$ = 40\,$\upmu$Jy beam$^{-1}$) and the intrinsic magnetic field direction of the emission. The green cross marks the optical centre of M82 and the blue ellipse the extent of the central optical disk.}
        \label{image_MF_18cm_22cm_lg_halo}
\end{figure}

Figures \ref{image_18cm_22cm_sm} and  \ref{image_MF_18cm_22cm_lg_sublr} show the peak polarised flux density recovered from the RM cubes of the combined $\uplambda$18\,cm and $\uplambda$22\,cm bands using the differential technique to remove the instrumental polarisation with resolutions of 15\farcs7$\times$14\farcs2 and 43\farcs1$\times$39\farcs6, respectively. Figure \ref{image_MF_18cm_22cm_lg_halo} shows the smoothed RM cube of the combined data with pixels in the centre of the galaxy showing instrumental polarisation blanked. Figures \ref{image_18cm_22cm_sm}-\ref{image_MF_18cm_22cm_lg_halo} were all derived from RM cubes out of RM=$\pm$ 600\,rad\,m$^{-2}$, but because of the way the differential technique works, Figs. \ref{image_18cm_22cm_sm} and \ref{image_MF_18cm_22cm_lg_sublr} do not include emission with RMs between $\pm$ 84\,rad\,m$^{-2}$. Figure \ref{image_MF_18cm_22cm_lg_halo} only shows emission where instrumental polarisation is negligible by comparison.

The polarised emission in Fig. \ref{image_18cm_22cm_sm} shows a patchy appearance with very low intensities close to the 5$\sigma$-detection limit. Two small regions are of special interest, one slightly north of the centre of M82 (Region 4) and one in the western disk (Region 5). The first one is in the area where the observations at $\uplambda$3\,cm and $\uplambda$6\,cm showed depolarisation. This can be a beam depolarisation effect. The observations at $\uplambda$18\,cm and $\uplambda$22\,cm might partly overcome this as long as the emitting layers are not positioned within the FWHM of the RMTF of each other. This might be the case if the observed structure is a superposition of the disk field and the field in the northern outflow cone. The retrieved intrinsic magnetic field vectors also fit to the VLA observations. The second region coincides well with the eastern part of the trough in Fig. \ref{image_RM_3cm_6cm}. The magnetic field vectors in this region are also mostly coincident with the ones in Figs. \ref{image_MF_3cm} and  \ref{image_MF_6cm} being disk parallel.

Figures \ref{image_MF_18cm_22cm_lg_sublr} and \ref{image_MF_18cm_22cm_lg_halo} show polarised emission in the halo of M82. Three patches towards the south of the galaxy at a distance of $\sim0.6$\,kpc (Region 6), $\sim1$\,kpc (Region 7), and $\sim1.5$\,kpc (Region 8) show a vertically directed magnetic field, which bends towards the east at a projected distance of $\sim1.6$\,kpc. In contrast to that, a large region towards the northwest is showing a magnetic field mainly parallel to the disk (Region 9). This patch shows a higher polarised intensity in regions with lower H$\upalpha$-emission.

The western patch in the southern halo (Region 7) matches well with a steeper gradient towards the halo in radio continuum total power measurements \citep{2013A&A...555A..23A}. A higher column density is also visible in kinematic studies in \ion{H}{i} \citep{1994Natur.372..530Y} and CO \citep{2002ApJ...580L..21W,2013PASJ...65...66S}, so that the regular magnetic field in this region could have been enhanced by compression due to the interaction.

\subsection{Rotation measures}

\begin{figure}
        \resizebox{\hsize}{!}{\includegraphics{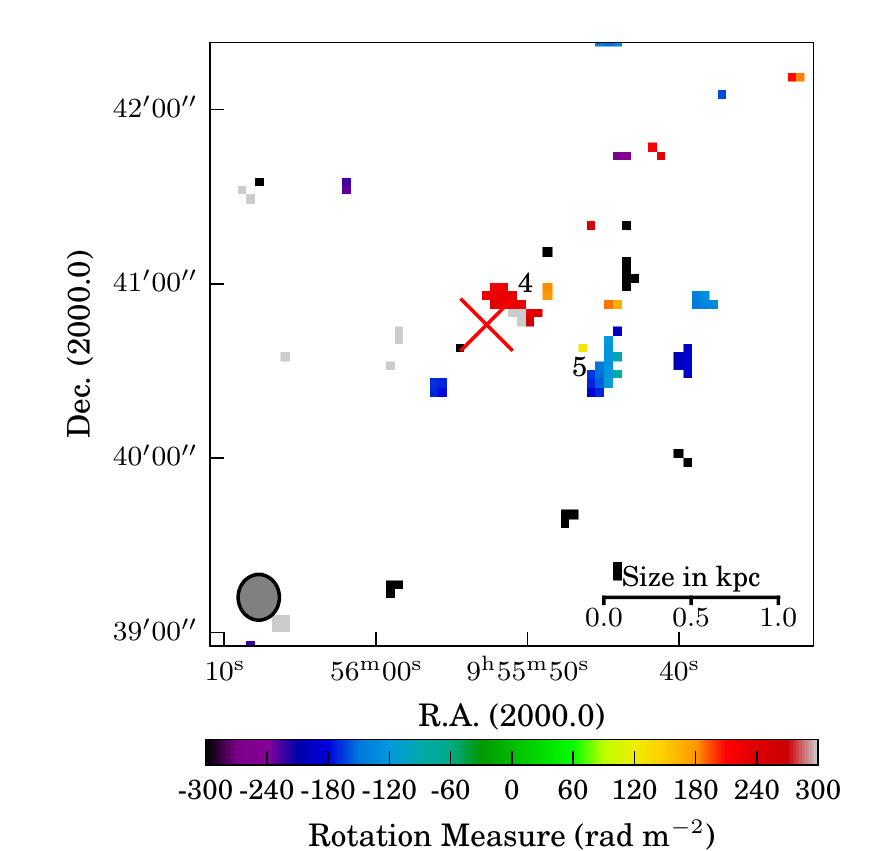}}
        \caption{RMs from the RM cube of the combined $\uplambda$18\,cm and $\uplambda$22\,cm data above a $5\sigma$ detection limit using the differential technique. In the lower left we show the size of the synthesised beam (15\farcs7$\times$14\farcs2). The galactic foreground RM of -19.65\,rad\,m$^{-2}$ was corrected for. Only $|RM|$ values between 84\,rad\,m$^{-2}$ and  600\,rad\,m$^{-2}$ were searched for in the RM cube on which the image was based. The red cross marks the optical centre of the galaxy.}
        \label{image_RM_18cm_22cm_sm}
\end{figure}

Figure \ref{image_RM_18cm_22cm_sm} shows the RMs derived from the high-resolution RM cube. The RMs north of the centre reach values as high as 300\,rad\,m$^{-2}$. The small curved structure towards the west shows RMs between $-60$\,rad\,m$^{-2}$ and $-180$\,rad\,m$^{-2}$.

\begin{figure}
        \resizebox{\hsize}{!}{\includegraphics{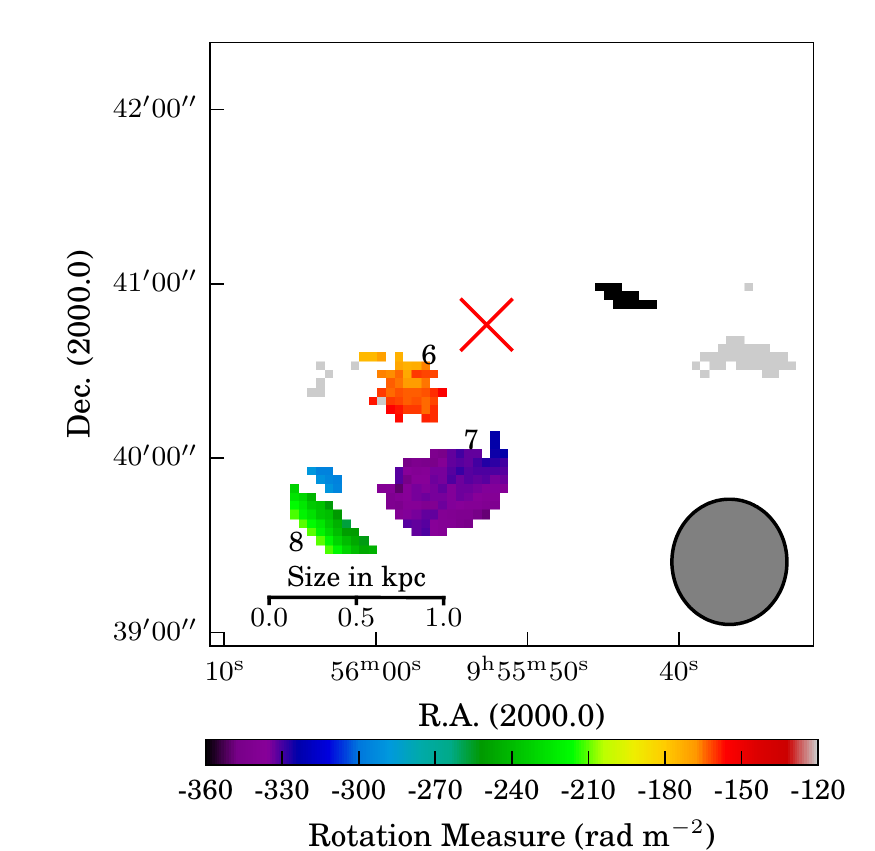}}
        \caption{RMs from the RM cube of the combined $\uplambda$18\,cm and $\uplambda$22\,cm data above a $5\sigma$ detection limit using the differential technique. In the lower right we show the size of the synthesised beam (43\farcs1$\times$39\farcs6). The galactic foreground RM of -19.65\,rad\,m$^{-2}$ was corrected for. Only $|RM|$ values between 84\,rad\,m$^{-2}$ and  600\,rad\,m$^{-2}$ were searched for in the RM cube on which the image was based. The red cross marks the optical centre of the galaxy.}
        \label{image_RM_18cm_22cm_lg_sublr}
\end{figure}

\begin{figure}
        \resizebox{\hsize}{!}{\includegraphics{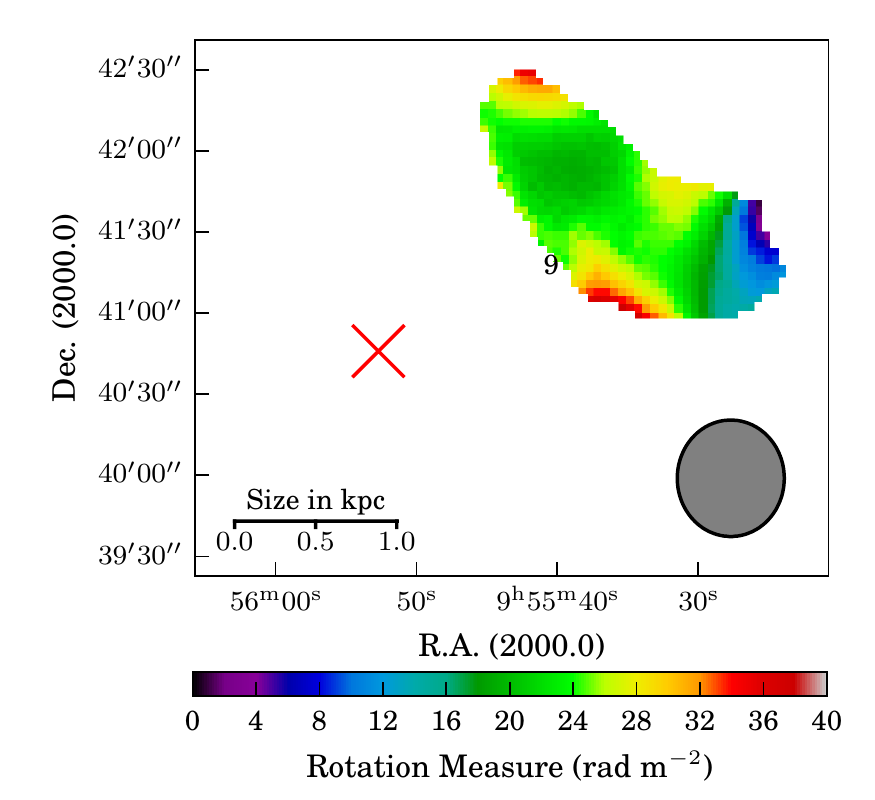}}
        \caption{RMs from the RM cube of the combined $\uplambda$18\,cm and $\uplambda$22\,cm data above a $5\sigma$ detection limit. The polarised instrumental signal at the centre of the galaxy was blanked. In the lower left we show the size of the synthesised beam (43\farcs1$\times$39\farcs6). The galactic foreground RM of -19.65\,rad\,m$^{-2}$ was corrected for. The red cross marks the optical centre of the galaxy.}
        \label{image_RM_18cm_22cm_lg_halo}
\end{figure}

The low-resolution RM maps (Figs. 9 and 10) show a clear asymmetry between the northern halo with positive RMs between 10\,rad\,m$^{-2}$ and 40\,rad\,m$^{-2}$ and the southern halo with RMs between -100\,rad\,m$^{-2}$ and -350\,rad\,m$^{-2}$. 

\section{Magnetic-field structure}
\label{text_magfield_struc}

The literature on M82 describes a poloidal magnetic-field morphology, where the magnetic-field lines are directed along the outflow direction of the superwind \citep{1994A&A...282..724R}. This was already proposed by \citet{1992A&A...256...10R} from a resolved study of the synchrotron halo of M82 revealing filaments and gaps, which were supposed to be created by a convecting magnetic field driven by the violent star formation in the centre along the outflow.

Our observations, adding another frequency at 1.4\,GHz, together with extended knowledge from other wavelengths and detailed analysis of gas compositions and motions in other literature, let us partly confirm this poloidal morphology for the southern halo, but we also detect magnetic-field structures indicating a more complex morphology, including a bar structure in the centre and hints for an X-shaped magnetic field in the halo of M82.

In the special case of M82 \citet{1989A&A...217...99L} proposed a ring current, which generates a poloidal magnetic field in the centre of M82. This idea was adopted by \citet{1994A&A...282..724R} and the same magnetic-field morphology was used by \citet{2000AJ....120.2920J} to explain their radio continuum and infrared observations, respectively. This idea could explain the slightly poloidal field in regions 3 and 4, but not the largely disk-parallel field on the western side (region 2).

\begin{figure}
        \resizebox{\hsize}{!}{\includegraphics{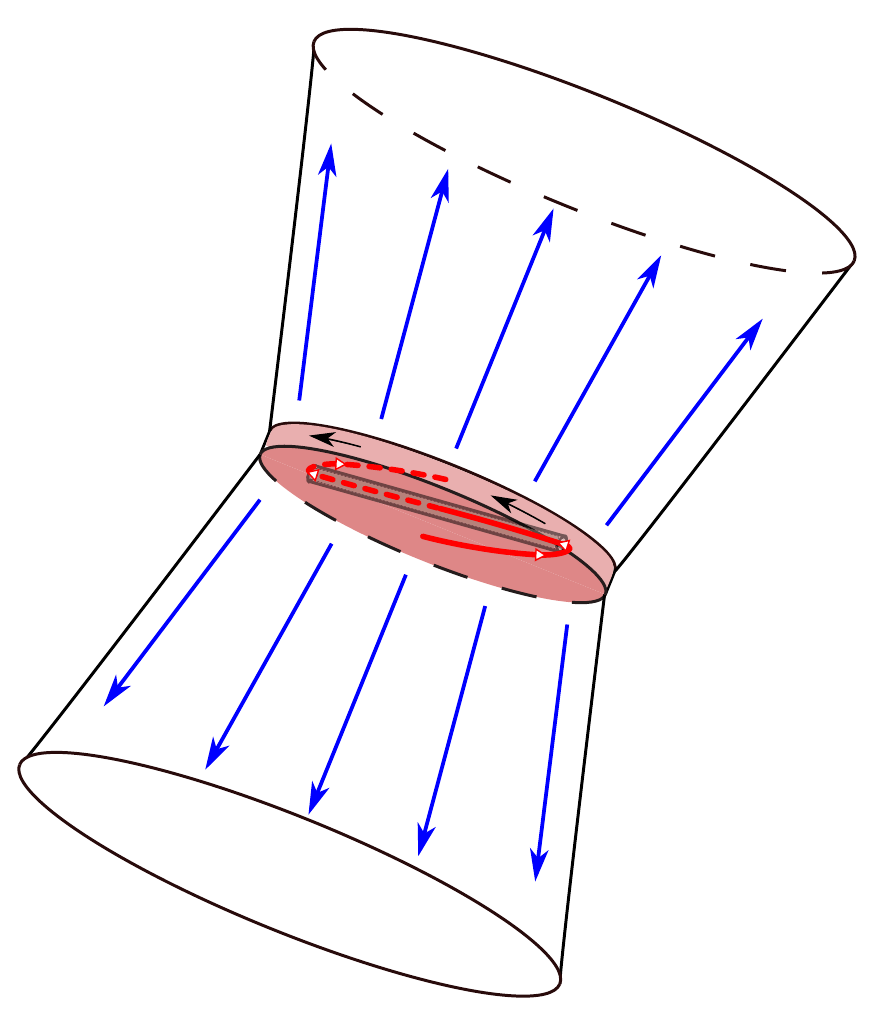}}
        \caption{Sketch of the magnetic field morphology in M82 showing our view of the galaxy. The grey cylinder represents the bar with the magnetic field as a red line. The black arrows on the disk show its rotation. The arrows on the red line show the direction of the magnetic field; the dashed part of the red line indicates that part of the magnetic field which is not observed. The direction of the outflow is shown with blue arrows.}
        \label{image_sketch}
\end{figure}

One possibility to explain the feature in region 2 would be an asymmetric structure in the galaxy, more precisely, by a magnetic field aligned with a bar (see Fig. \ref{image_sketch} for a sketch of the morphology). 

RMs determined using
\begin{equation}
        \text{RM} = 0.81 \int_{0}^{\infty} n_e B_{||} \text{d}L,
        \label{Eq:RM}
\end{equation}
with the regular magnetic-field strength $B_{||}$ in $\upmu$G; the thermal electron density $n_e$ in cm$^{-3}$ allowed us to determine whether a field is pointing towards the observer or away from them. Therefore Fig. \ref{image_RM_3cm_6cm} indicates a field which points away from the observer on the western side and therefore towards the centre of M82 if the field is aligned with the bar.

The findings of a bar structure by \citet{1991ApJ...369..135T}, \citet{1995ApJ...439..163A}, and \citet{2000MNRAS.316...33W} in the infrared, [Ne II], and \ion{H}{i} absorption measurements, respectively, indicate such a scenario as well. \citet{2005ApJ...628L..33M}, using infrared observations, even determined the length of the bar to be $\sim$1\,kpc and also detected spiral arms after subtracting an exponential profile for the disk. The trough at a projected distance of $\sim$500\,pc from the centre of the galaxy in the RM map (Fig. \ref{image_RM_3cm_6cm}) is consistent with this morphology. \citet{2002A&A...391...83B} showed that magnetic fields can be aligned with bars in galaxies and also with the arm or interarm regions \citep{2011MNRAS.412.2396F}. The smooth transition to positive RM values further west can be explained by the spiral arm beginning from the tip of the bar turning towards the backwards part of the galaxy away from the observer at a high pitch angle as has been observed by \citet{2005ApJ...628L..33M}. The positions of the detected features fit the former observations. This morphology would indicate a bisymmetric spiral field \citep{1990IAUS..140..187K} for the inner disk (and bar) of M82. 

The missing polarised emission in the east could be caused by depolarisation due to the strong turbulent magnetic field in front of the bar or in the bar itself. In addition, \citet{2000MNRAS.316...33W} reported a truncation of the bar towards the east. Distinguishing between these two scenarios is not possible with our data.

The magnetic-field morphology in the southern part of the galaxy is mainly pointing away from the disk, but regions 6 to 8 show that the magnetic field bends towards the disk. This could well be a regular magnetic field which had its origin in the disk and was transported out into the halo by the strong outflow causing a loop-like structure in the halo. Indications for such large-scale loop-like structures have already been found for the nearby starburst galaxy NGC1569 \citep{2010ApJ...712..536K}.

The disk-parallel magnetic field in the northern halo (Region 9) cannot be explained by a simple transport along the outflow. The small positive RMs in this region could indicate that the polarised structure belongs to the near side of the northern outflow cone. Assuming a cylindrical or plume-like geometry \citep{2010A&A...514A..14K} and a magnetic field pointing away from the disk of the galaxy on both sides would most likely result in RMs of different signs. The smaller absolute RMs of the northern halo in contrast to the southern one can have two different origins. Assuming a symmetry of the morphology of the outflow cones, where the southern cone points towards the observer and the northern away and considering the opening angle of the cones would result in a smaller projected magnetic field component in the north towards the observer giving smaller RMs. The second option would be a higher magnetic-field strength and/or electron density towards the south, which would not be unexpected due to the proximity of the other group galaxies towards this direction.

\section{Magnetic-field strength}
\label{text_magfieldstr}

Assuming equipartition between the energy of the magnetic field and the cosmic ray electrons allows us to calculate the regular and turbulent magnetic-field strengths in all marked regions using the equations by \citet{2005AN....326..414B}. A proton to electron ratio $K=100$ was assumed in all cases. \citet{2013MNRAS.430.3171L} showed that the equations from \citet{2005AN....326..414B} give reasonable results even for starburst galaxies if secondary particles are included. In this case only small corrections need to be applied to the equations from \citet{2005AN....326..414B}.

For all calculations we assumed only one emitting component for each line of sight in the RM cubes. We used the average flux densities of the total power maps in \citet{2013A&A...555A..23A} and the polarised intensity maps in this paper to derive the fractional polarisation of the individual regions. Only pixels above the specified thresholds (the lowest intensity contours in the used maps) were considered to be reliable. Spectral index values were also taken from \citet{2013A&A...555A..23A} using the same method. The equation from \citet{2005AN....326..414B} can only be solved for spectral indices $\alpha<-0.5$, where $S_\nu \propto \nu^\alpha$. Some regions in the $\uplambda$22\,cm maps show higher values due to absorption effects. In those cases we used the average value for the centre of $\alpha=-0.62$.

\begin{figure}
        \resizebox{\hsize}{!}{\includegraphics{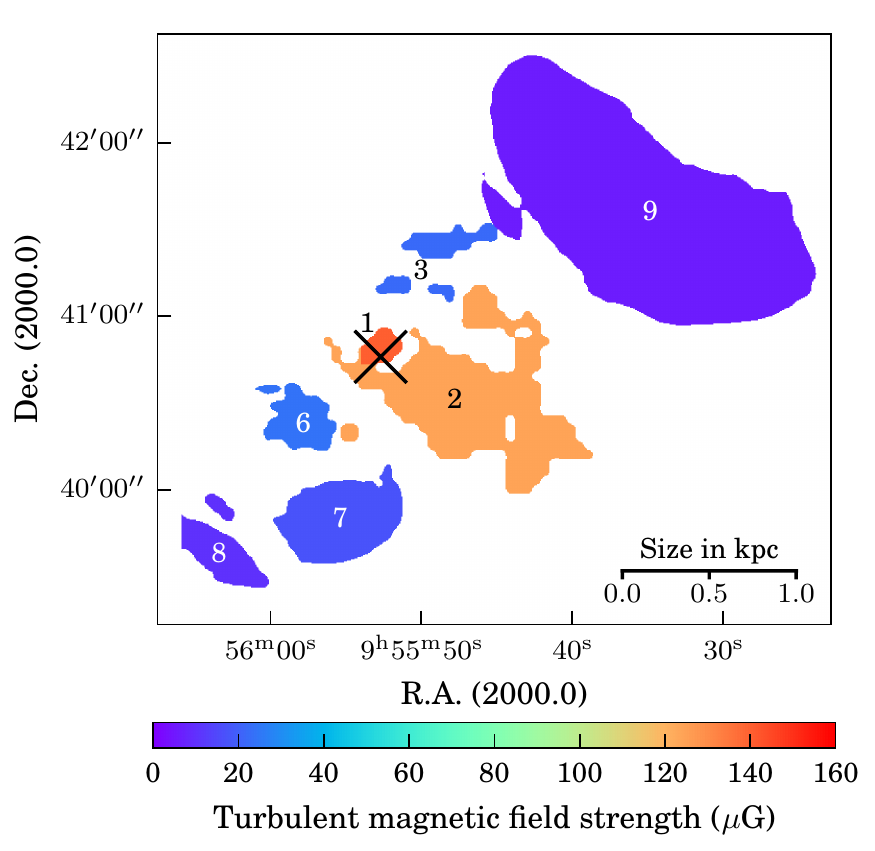}}
        \caption{Turbulent magnetic field strength taken from Tab. \ref{table_Bfield}. The corresponding regions are indicated by the numbers. The black cross marks the optical centre of the galaxy.}
        \label{image_Bfield_turb}
\end{figure}

\begin{figure}
        \resizebox{\hsize}{!}{\includegraphics{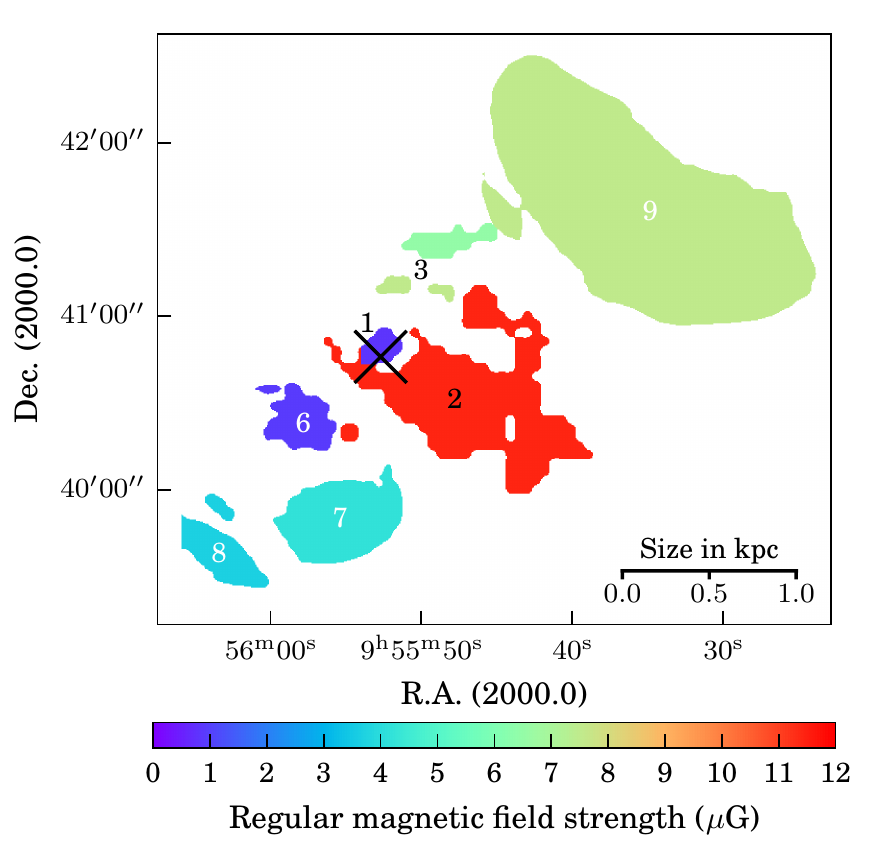}}
        \caption{Regular magnetic field strength taken from Tab. \ref{table_Bfield}. The corresponding regions are indicated by the numbers. The black cross marks the optical centre of the galaxy.}
        \label{image_Bfield_reg}
\end{figure}

To calculate the regular magnetic-field strength, we need the inclination angle of the magnetic field with respect to the plane of the sky and the path length of the emitting medium. In the special case of the bar, this is directly connected to its orientation and extend. \citet{1991ApJ...369..135T} reported a value for the inclination of $i=22^\circ$ from their near-IR measurements. Since the magnetic field in other barred galaxies was reported to be regular with a small offset in the upstream direction \citep{2005A&A...444..739B} and not directly on top of the bar, we used $i=30^\circ$ as an assumption. The approximate size of the emitting region was derived from the width of the polarised structure in our $\uplambda$3\,cm and $\uplambda$6\,cm maps in direction of the minor axis leading to $z\approx250$\,pc. For the halo magnetic field, we assume the same inclination as the disk of the galaxy of $i=11^\circ$. All calculated values are listed in Tab. \ref{table_Bfield} and the turbulent and regular magnetic field strength are illustrated in Figs. \ref{image_Bfield_turb} and \ref{image_Bfield_reg}, respectively.

\begin{table*}
        \caption{Parameters used for calculating the turbulent B$_{turb}$ and regular magnetic field strengths B$_{reg}$}
        \label{table_Bfield}
        \centering
        \begin{tabular}{c c c c c c c c c c c}
                \hline
                \hline
                Region & $\nu$ (GHz) & TI (mJy) & PI (mJy) & P (\%) & $\alpha$ & $i$ ($^\circ$) & $z$ (pc) & $B_{turb}$ ($\upmu$G) & $B_{reg}$ ($\upmu$G) & Location \\
                \hline
                1 & 8.44 & 488 & 0.125 & 0.026 & -0.60 & 11 & 900 & 140 & 0.8 & centre \\
                2 & 8.44 & 98 & 0.255 & 0.258 & -1.44 & 30 & 250 & 124 & 11.4 & bar \\
                3 & 8.44 & 0.89 & 0.11 & 12.00 & -0.88 & 11 & 900 & 22 & 7.5 & halo \\
                1 & 4.86 & 678 & 0.275 & 0.041 & -0.60 & 11 & 900 & 140 & 1.1 & centre \\
                2 & 4.86 & 144 & 0.224 & 0.155 & -1.44 & 30 & 250 & 113 & 8.4 & bar \\
                3 & 4.86 & 1.43 & 0.133 & 9.28 & -0.88 & 11 & 900 & 22 & 6.5 & halo \\
                4 & 1.54 & 1249 & 0.076 & 0.006 & -0.62 (-0.14) & 11 & 900 & 117 & 0.3 & centre \\
                5 & 1.54 & 89 & 0.049 & 0.056 & -0.84 & 30 & 250 & 67 & 1.3 & bar \\
                6 & 1.54 & 33 & 0.014 & 0.041 & -1.40 & 11 & 900 & 24 & 0.9 & halo \\
                7 & 1.54 & 5.37 & 0.159 & 2.90 & -1.68 & 11 & 900 & 17 & 4.2 & halo \\
                8 & 1.54 & 1.40 & 0.132 & 9.44 & -1.26 & 11 & 900 & 10 & 3.7 & halo \\
                9 & 1.54 & 0.47 & 0.243 & 51.4 & -1.61 & 11 & 900 & 5.8 & 7.5 & halo \\
                \hline
        \end{tabular}
        \tablefoot{Regions are numbered as in the images. $\nu$ is the average observing frequency, TI, P, and $\alpha$ are the average total intensity, polarised intensity, and spectral indices over the integrated area, FP is the average fractional polarisation calculated from P/TP, $i$ the inclination of the regular magnetic field towards the observer, and $z$ the assumed path length of the emitting medium. The most likely location of the detected structure is given in the last column.}
\end{table*}

\subsection{Centre}

The turbulent magnetic-field strength B$_{turb}$ in the centre shows very high values between 117\,$\upmu$G and 140\,$\upmu$G. Such high field strengths are most likely the result of a superposition at least two different phases of the magnetised medium, the strong magnetic fields in the mG regime in compact star-forming regions and the diffuse component surrounding them \citep{2013A&A...555A..23A}. If we assume a similar turbulent magnetic-field strength as in other starburst galaxies of B$_{turb}=20\,\upmu$G \citep{2011MNRAS.412.2396F, 2007A&A...470..539B} for the diffuse component, we reach filling factors of not more than 17\% for the compact star-forming regions. The aforementioned values did not account for the different phases of the magnetised medium, meaning that their turbulent magnetic field strength is an upper limit for the diffuse component.

In contrast to that, the strength of the large-scale magnetic field B$_{reg}$ shows low values of 0.8\,$\upmu$G ($\uplambda$3\,cm), 1.1\,$\upmu$G ($\uplambda$6\,cm), and 0.3\,$\upmu$G ($\uplambda$22\,cm). The value at $\uplambda$22\,cm is most likely influenced by depolarisation effects, while the values at $\uplambda$3\,cm and $\uplambda$6\,cm nearly show the same field strengths and fractional polarisations, indicating that they suffer less from depolarisation, as expected.

The traced polarised flux intensities could be the result of ordered magnetic fields in giant molecular clouds. Such magnetic fields can be tied to the mean magnetic field direction as has been observed for M33 \citep{2011Natur.479..499L}.

Using Eq. \ref{Eq:RM} with $B_{||}=0.8$\,$\upmu$G, RM=600\,rad\,m$^{-2}$ from Fig. \ref{image_RM_3cm_6cm}, and $n_e=18.7$\,cm$^{-3}$ from thermal absorption measurements \citep{2013A&A...555A..23A}, we are able to calculate the typical coherence length to $\langle L \rangle=50$\,pc. This is similar to the sizes of giant molecular clouds.

It has to be mentioned that the fractional polarisation values for these calculations are below the threshold for the instrumental contribution (see Sect. \ref{text_instpol_polbias}), so that they could be contaminated by instrumental effects. The limit was derived from the noise in the calibrator measurement where no signal was found, so that the instrumental contamination is just an upper limit.

\subsection{Bar}

The regular field in this region is much stronger than the one in the centre with 11.4\,$\upmu$G derived from observations at $\uplambda$3\,cm. \citet{1999Natur.397..324B} derived similar field strengths in NGC1097. Again the much lower value at $\uplambda$22\,cm of 1.3\,$\upmu$G is most likely influenced by depolarisation effects. The turbulent field strengths are similar to the ones in the central region and therefore might be a superposition of the field in the bar and the one in the central region since most of the bar is lying inside the central starburst region.

\subsection{Halo and outflow cones}

The magnetic field strengths for the halo regions show a decrease of the turbulent magnetic field strength and an increase of the regular one towards the outer parts of the halo. While the average turbulent field strengths for the northern and southern halo are similar and on the order of 17$\pm7$\,$\upmu$G, the regular field strength in the southern halo (0.9-3.7\,$\upmu$G) reaches only $\approx50$\% of the northern one (7.5\,$\upmu$G) and less than $\approx2$\% of the strength of the turbulent field in the centre of M82. Region 9 is of special interest since it is showing a region where the ordered field dominates the turbulent one. This is known from normal star-forming spiral galaxies and could indicate that strong ordered magnetic fields in the surroundings exist even for starburst galaxies.

\subsection{Feeding the central starburst via magnetic braking}
\label{text_magnetic_braking}

An ionised ring with a radius of $\approx25$\,pc \citep{2000MNRAS.316...33W} and a ring of neutral gas with a radius of $\approx500$\,pc \citep{1993ApJ...411L..17Y} are two of the features in the central part of M82. Neutral gas in the outer ring can lose angular momentum, and get ionised when it falls towards the centre of M82 \citep{1994MNRAS.270..703J,2000MNRAS.316...33W}. Fuelling the nuclear starburst of M82, where stars form at a rate of 13\,M$_\odot$yr$^{-1}$ \citep{2003ApJ...599..193F} is still difficult to explain based on only gravitational forces, as has been discussed by several authors \citep[e.g.][]{1995ApJ...449..508P,2004ASSL..319..189K}.

One of the ideas proposed by \citet{1999Natur.397..324B} for the barred galaxy NGC\,1097 with a lower star-formation rate of 5\,M$_\odot$yr$^{-1}$ is the active driving of the inflow by magnetic forces. In \citet{2013A&A...555A..23A} we already showed that the magnetic field is tightly coupled to the warm ionised medium in M82. Since the centre of M82 is mostly ionised, containing both a strong large-scale magnetic field most likely aligned with the bar with a large pitch angle and a turbulent field on smaller scales as well as a circumnuclear ring, we can assume the same mechanism as described in \citet{1999Natur.397..324B}. The described mechanism estimates the Lorentz force on the gas in a magnetised circumnuclear ring where the turbulent field is contributing most to the loss of angular momentum. Using the equation for the mass inflow rate $\dot{M}$ from \citet{2005A&A...444..739B}

\begin{equation}
        \dot{M} = \frac{h}{\Omega}\left( \overline{B_{turb}^2} - \frac{1}{2}\,\overline{B_{reg}}^2 \sin (2p)\right),
        \label{Eq:Acc}
\end{equation}

\noindent with $h=0.1$\,kpc being the disc scale height, where we used an average value from the fits in \citet{2013A&A...555A..23A}, $\Omega=217\,$km\,s$^{-1}$kpc$^{-1}$ is the angular velocity of the bar \citep{2000MNRAS.316...33W}, $B_{turb}=124\,\upmu$G is the turbulent field strength, $B_{reg}=11.4\,\upmu$G is the regular field strength and $p=45^\circ$ is the pitch angle of the magnetic field lines; we obtain $\dot{M}=7.1\,$M$_\odot$yr$^{-1}$, which is therefore a significant contribution to the star-formation rate. Here we used the maximum value for $\sin(2p)$. The dominant factor in this equation in our case is the turbulent field. The regular field strength only reaches 7\%-9\% of the turbulent one. Therefore even strong deviations of $p$ from our assumed value would not affect the mass inflow rate significantly.

\section{Depolarisation}
\label{text_depol}

\begin{figure}
        \resizebox{\hsize}{!}{\includegraphics{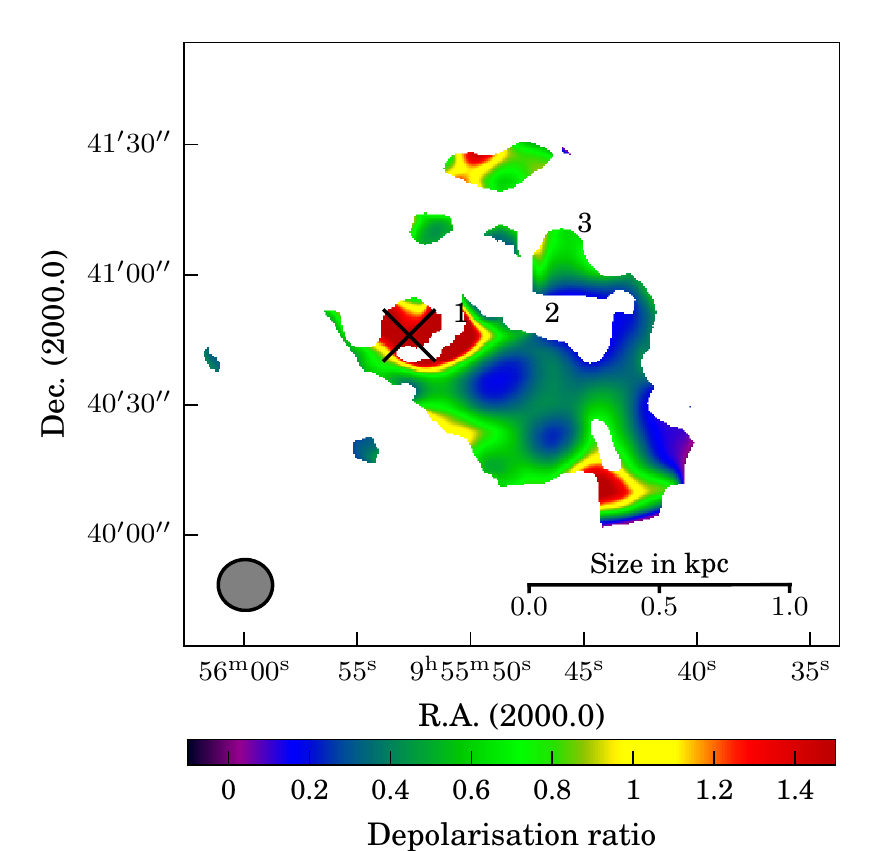}}
        \caption{Depolarisation ratio between $\uplambda$3\,cm and $\uplambda$6\,cm. In the lower left we show the size of the synthesised beam (12\farcs5$\times$11\farcs7). The black cross marks the optical centre of the galaxy.}
        \label{image_DP3cm6cm}
\end{figure}

The ratio of the fractional polarisation between $\uplambda3\,$cm and $\uplambda6\,$cm is shown in Fig. \ref{image_DP3cm6cm}. We did not account for changes of the polarisation spectrum due to spectral index effects since our total intensity values are influenced by thermal absorption \citep{2013A&A...555A..23A}, and therefore we were not able to reconstruct the synchrotron spectral index. The resulting map shows the depolarisation ratio $DP$ between $\uplambda$3\,cm and $\uplambda$6\,cm with a value of 1 representing no depolarisation and a value of 0 representing complete depolarisation.

Inverted depolarisation ratios of up to 1.4 in the centre of the galaxy are visible. Such inverted polarisation ratios are known from spatially resolved studies of other nearby galaxies \citep[e.g.][]{1992A&A...265..417H,1991A&A...251...15B} and can be explained by either a twisted helical field \citep{1998MNRAS.299..189S} or a medium with at least two emitting components. Within a twisted helical field configuration the intrinsic polarisation angle varies along the minor axis. If polarisation angles of different contributing and superimposing components are equally changing, the observed net polarisation would not be depolarised at all. Since Faraday Rotation depends on the observed wavelength, this can only occur for a certain wavelength range. The same phenomenon is possible in the case of two emitting layers where the magnetic field orientation changes abruptly.

\citet{2011A&A...535A..79H} already detected a helical magnetic field in the innermost outflow region of the starbursting galaxy NGC253. One of the signs for such a configuration is the rapid switch of the sign of the RMs along this structure, which is also visible in our map (Fig. \ref{image_RM_3cm_6cm}). We therefore favour the same morphology for the central region of M82. Using the equation

\begin{equation}
        \lambda^2 < \frac{\Delta\Phi_0}{RM},
        \label{Eq:helical}
\end{equation}

\noindent we can calculate at which wavelength $DP=1$ from $\Delta\Phi_0$, if we know the change of the magnetic field across our observed region $\Delta\Phi_0$ in rad. Using $\Delta\Phi_0=\pi$ and $RM=600$\,rad\,m$^{-2}$ to acquire an upper limit for the highest possible wavelength where an increased polarisation occurs already gives us $\lambda\leq7.2$\,cm. This value can easily be smaller if we assume that the helix does not completely twist by $180^\circ$ inside the investigated area.

Fig. \ref{image_DP3cm6cm} shows that the depolarisation ratio decreases to values between 0.2 and 0.6 towards the west of the central region with two nearly circular structures where $DP$ reaches minimal values. Those have an offset towards the east and west of the trough in Fig. \ref{image_RM_3cm_6cm}.

Apart from the instrumental depolarisation effects like bandwidth or beam depolarisation discussed above, several depolarisation effects caused by the morphology of the emitting or intervening magnetic field can be considered. For the following calculations \citep{1966MNRAS.133...67B} we use the assumption of a uniform slab with uniform synchrotron emission and Faraday rotation along the line of sight. More complicated models were proposed by \citet{1998MNRAS.299..189S} which extended the equations by \citet{1966MNRAS.133...67B}. Due to our sparse frequency coverage (only 3-4 data points over the whole frequency range) our analysis is limited to the above-mentioned simple morphology. Analysis of more complex morphologies would need advanced techniques like Q-U-fitting routines or modelling (See \citep[e.g.][]{2015MNRAS.450.3579S}), which are only applicable in cases of good frequency coverage and high signal-to-noise ratios.

Differential Faraday Rotation (DFR) occurs in the case of an extended emitting and rotating region. Because of the wavelength-dependent rotation of the received electric field vectors, emission from the backside of the emitting structure is therefore rotated more than that from the front. This leads to a sinc behaviour of the received fractional polarised flux density over wavelength and therefore to a complete cancellation for certain combinations of wavelengths and RMs. The equation describing this dependence is:
\begin{equation}
        DP_{DFR} = \frac{|\sin (2\,\text{RM}\,\lambda^2)|}{|2\,\text{RM}\,\lambda^2|}
        \label{Eq:DFR}
.\end{equation}
Since this depolarisation effect is solely dependent on the wavelength $\lambda$ and the RM, only the regular magnetic field component can cause it. In contrast to this, the turbulent magnetic field component can cause Faraday Dispersion, which is dependent on the turbulent cell size  $d$ in pc, the path-length $L$ in pc, the average electron density $\langle n_e \rangle$ in cm$^{-3}$ along the depolarising layer, the average turbulent magnetic field strength $\langle B_{turb} \rangle$ in the depolarising layer, and the filling factor of the turbulent cells causing the depolarisation $f$:
\begin{equation}
        \sigma^2 = (0.81\langle n_e \rangle \langle B_{turb}\rangle)^2 Ld / f.
        \label{Eq:RMdisp}
\end{equation}
We can distinguish between two different Faraday Dispersion effects; External Faraday Dispersion (EFD) and Internal Faraday Dispersion (IFD). In EFD the turbulent magnetic field lies in front of the source and the corresponding depolarisation can be calculated using
\begin{equation}
        DP_{EFD} = \exp(-2\,\sigma^2\lambda^4)
        \label{Eq:EFD}
.\end{equation}
In the case of IFD the depolarising layer is situated inside the source and the above equation becomes
\begin{equation}
        DP_{IFD} = \frac{1 - \exp(-2\,\sigma^2\lambda^4)}{2\,\sigma^2\lambda^4}
        \label{Eq:IFD}
.\end{equation}
In addition, \citet{2008A&A...487..865R} proposed a model of a Faraday depolarising medium, which is not fully filling the line of sight with turbulent cells leading to less depolarisation especially for high-resolution observations at long wavelengths. A simple model using this assumption would result in a constant offset of all data points over the whole frequency range. Unfortunately, reliable fits would again need a better frequency coverage.

\begin{figure}
        \resizebox{\hsize}{!}{\includegraphics{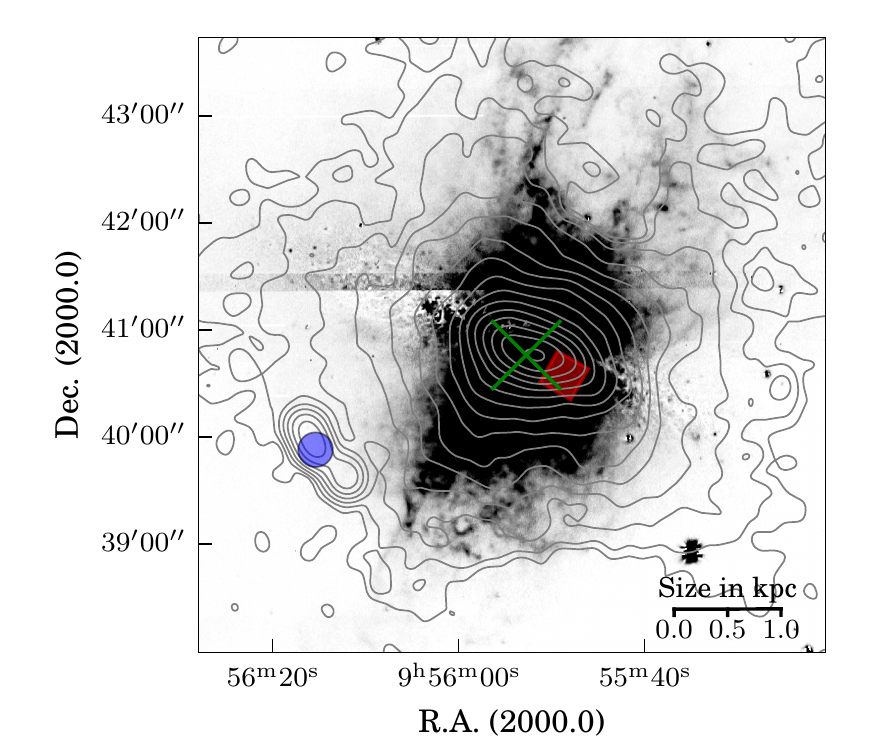}}
        \caption{Total power radio continuum contours at $\uplambda$22\,cm from the WSRT observations in \citet{2013A&A...555A..23A} on a H$\upalpha$ image from the WIYN 3.5m Observatory. Contours start at a 3$\sigma$ level of 90$\upmu$Jy/beam and increase in powers of 2. The green cross marks the optical centre of M82, the red box the bar region, and the blue circle the position of the background point source.}
        \label{plot_pos_DP_sources}
\end{figure}

\subsection{Bar}

\begin{figure}
        \resizebox{\hsize}{!}{\includegraphics{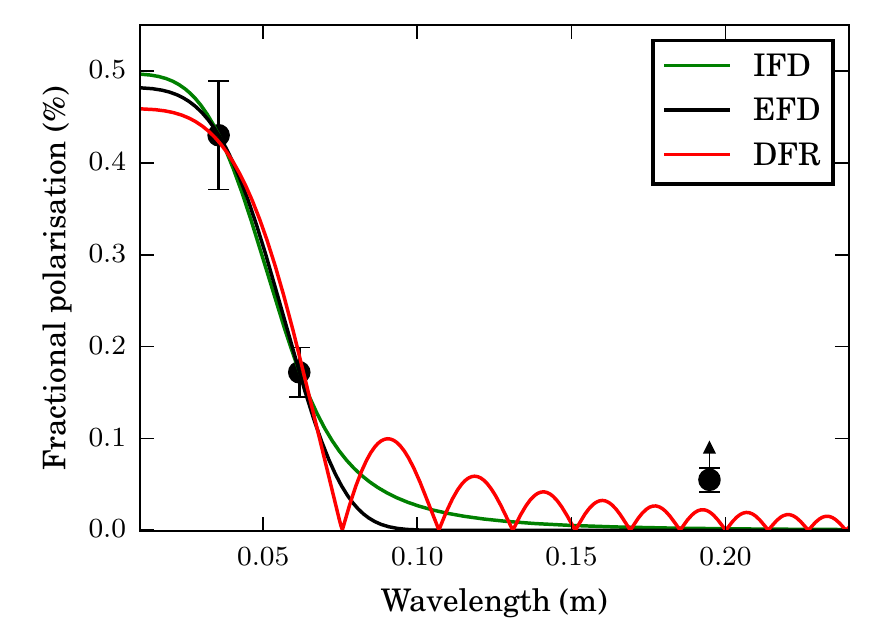}}
        \caption{Wavelength plotted against the fractional polarisation in the bar region of M82. The lines show the best fits for IFD, EFD, and DFR.}
        \label{plot_DP_bar}
\end{figure}

Polarisation in the bar region was detected in all three wavelength regimes used in this paper ($\uplambda$3\,cm, $\uplambda$6\,cm, $\uplambda$18/22\,cm). We therefore can use the polarised flux densities of this region to learn more about the morphology of the magnetic field. We integrated the polarised and total flux densities over an area of 20\arcsec$\times$20\arcsec centred on the tip of the bar structure to calculate the fractional polarisation for each individual wavelength regime. Due to the low level of polarisation in the combined $\uplambda$18/22\,cm RM cube, we were not able to detect the bar in the individual $\uplambda$18\,cm and $\uplambda$22\,cm measurements. Additionally, the differential technique which we used to suppress instrumental polarisation prohibits a reliable integration over the RM axis, so that we only used the strongest recoverable component. Therefore the specified degree of polarisation is only a lower limit. Least-squares fitting was then used leaving only the intrinsic degree of polarisation $p_i$ and the Faraday Dispersion $\sigma$ or the RM in case of DFR as free parameters. The three different results are shown in Fig. \ref{plot_DP_bar}.

All fits give a similar value for the intrinsic degree of polarisation (IFD: $0.50\pm0.05$\,\%, EFD: $0.48\pm0.07$\,\%, DFR: $0.46\pm0.08$\,\%) since these are dominated by the $\uplambda$3\,cm datapoint. IFD gives a value of $\sigma_{IFD}=303.86\pm 76.06$\,m$^{-2}$ while EFD gives $\sigma_{EFD}=188.52\pm 35.42$\,m$^{-2}$. All three effects and also more complicated models \citep{2015MNRAS.450.3579S} show a rapid decrease of the fractional polarisation towards longer wavelengths.

Using the simple model, all effects can explain the depolarisation between $\uplambda$3\,cm and $\uplambda$6\,cm equally well. To explain the value at $\uplambda$18/22\,cm we need to include the measurements of the RM at this wavelength (Fig. \ref{image_RM_18cm_22cm_lg_sublr}). In the case of DFR, we would never be able to measure $|RM|$ $\geq$ 40\,rad\,m$^{-2}$ from the whole polarised emission of the bar. Emission from the backside would already have been cancelled out due to the superposition with vectors in front of it. We therefore conclude that the dominant depolarisation effect is caused by the turbulent field. We can imagine two different scenarios: The first one being a regular magnetic field in the bar in the background, where the emission is depolarised due to a turbulent magnetic field in front of it (EFD). The second scenario would be a colocation of the ionised gas and the turbulent field (IFD), where depolarisation occurs in the bar itself.

Since our measured RMs of -180\,rad\,m$^{-2}$ are close to the result for EFD, we favour the former scenario slightly, but are unable to exclude  the latter. The detection of polarised emission at $\uplambda$18/22\,cm leaves us with the idea of a perforated medium, where emission for some lines-of-sight is not depolarised at all since they never encounter a turbulent cell. Those cells most likely follow the distribution of dense star-forming regions or giant molecular clouds, which only occupy between $2\%$ \citep{2013A&A...555A..23A} and $17\%$ (this publication) of the volume.

\subsection{Background source}

A polarised background source was detected at all observed wavelengths in the southern halo. It lies at $\upalpha=09^\text{h} 56^\text{m} 15\fs39, \updelta=+69\degr 39\arcmin 52.9\arcsec$ with a projected distance of 2.5\,kpc from the centre of M82. This source was already detected in the NVSS RM catalogue (identifier NVSS\,095615+693952) by \citet{2009ApJ...702.1230T} with an integrated flux density of $17.8\pm1.0$\,mJy and no reliable polarised intensity ($-0.35\pm 0.52$\,mJy). Our maps (Fig. \ref{plot_pos_DP_sources}) show a double source at this position, most likely a double lobe radio galaxy. Polarisation was only detected in the western lobe, therefore we were able to fit Gaussians to the total power maps and use only the western component for our determination of the fractional polarisation.

\begin{figure}
        \resizebox{\hsize}{!}{\includegraphics{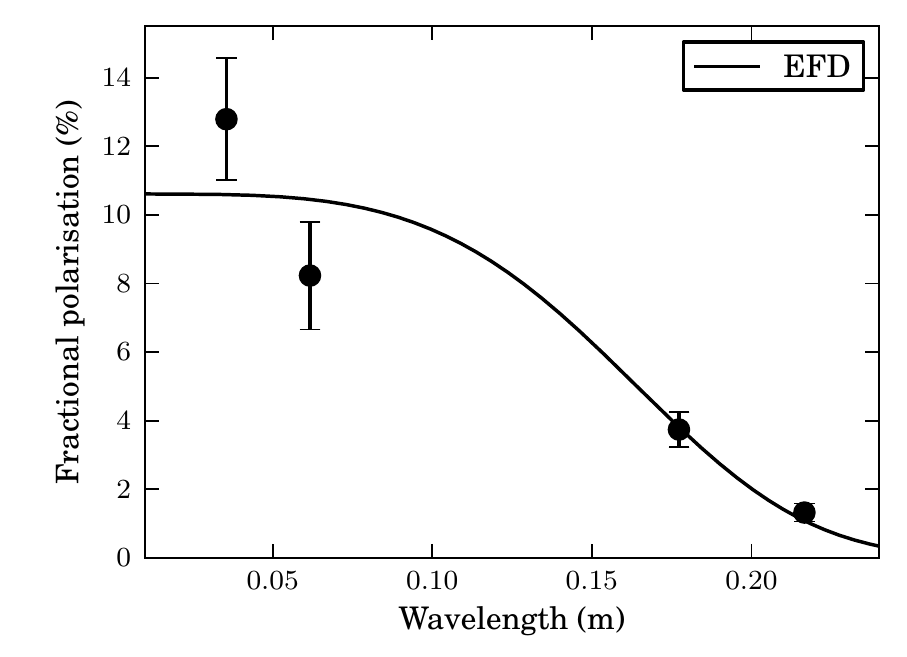}}
        \caption{Wavelength plotted against the fractional polarisation in the halo source. The black line shows the best fit for depolarisation by external Faraday dispersion (EFD) by the turbulent magnetic field in the halo of M82.}
        \label{plot_DP_halosource}
\end{figure}

The polarisation spectrum of the background radio source can be modelled using EFD. We only left the RM dispersion $\sigma$ and the intrinsic fractional polarisation $p_i$ as free parameters for the fit (Fig. \ref{plot_DP_halosource}) giving us $p_i=10.61\pm1.59$\,\% and $\sigma=22.77\pm5.67$\,rad\,m$^{-2}$. The depolarisation of the signal can occur all along the line of sight, but is
probably dominated by the background source itself and the halo of M82. If we assume that the depolarisation is dominated by the halo of M82, we can estimate the electron density in the halo of M82 using the definition of the RM dispersion (Eq. \ref{Eq:RMdisp}). Assuming $L=900$\,pc (the average diameter of the outflow cone), $d=100$\,pc, $f=1$, and $B_{turb}=10$\,$\upmu$G (Region 8 in Tbl. \ref{table_Bfield}) results in an average electron density of $\langle n_e \rangle\approx0.009$\,cm$^{-3}$. This is on the same order as the values in the Milky Way, which reach values between 0.007\,cm$^{-3}$ \citep{2012MNRAS.427..664S} and 0.002\,cm$^{-3}$ for a distance of 2.5\,kpc from the midplane. These two galaxies in fact host fundamentally different star-formation environments and gas kinematics. The similarity can be explained by a higher outflow momentum in the M82 superwind and therefore higher outflow velocities instead of higher electron densities in their halos. It must be noted that a certain fraction of the depolarisation might occur in the background source itself, meaning that the value for $\sigma$ is only an upper limit and therefore $n_e$ might be lower. 

In addition, we were able to measure the RM of the source using our $\uplambda$18/22\,cm data to $\text{RM}=-63.96\pm1.15$\,rad\,m$^{-2}$, or $\text{RM}=-44.31$\,rad\,m$^{-2}$ after correcting for the RM contribution by the Milky Way (Sect. \ref{Sect:18/22}). The negative sign confirms the morphology of a magnetic field pointing away from the observer on the southern side. With the value of the RM we are also able to calculate the regular magnetic field strength using our value of $\langle n_e \rangle\approx0.009$\,cm$^{-3}$ and Eq. \ref{Eq:RM} resulting in $B_{ord}\approx6.5$\,$\upmu$G. This is similar to the value we derived via the energy equipartition calculations.

\section{Discussion}
\label{text_discussion}

The morphology of the magnetic field and the warm ionised gas are closely coupled. We have shown that this has a significant impact on the overall spectrum in total flux density, because of thermal absorption effects causing a flattening of the spectrum in the starburst region towards longer wavelengths \citep{2013A&A...555A..23A}, as well as in polarised flux density, because of the different depolarisation phenomena discussed in Sect. \ref{text_depol}. The reason is that starburst galaxies seem to host two well separated phases of the warm ionised medium; a dense one, which is confined to dense molecular clouds where the actual intense star formation happens and which might be the dominant contribution to the warm ionised gas fraction; and a rarified one filling the space in between those regions. The result for the overall total flux density spectrum are two thermal absorption features \citep{2013MNRAS.431.3003L}.

This kind of gas distribution also has a direct impact on the magnetic field morphology and its associated polarisation spectra since magnetic fields are coupled to the ionised gas. Whether or not the magnetic fields dominate the gas dynamics or vice versa depends on the strength and orientation of the field compared to the density, velocity, and direction of movement of the gas \citep{2005A&A...436..585D,1993ApJ...409..663M}.

In case of a polarised source in the background of the galaxy or in the galaxy itself, the magnetic field inside the star-forming regions has a much smaller scale length than the angular resolution one can achieve with radio interferometers causing it to appear turbulent and depolarise the emission. Since not all emission crosses a turbulent cell it is still possible to recover polarised emission even at higher wavelengths.

For our observations the polarised background structure is a magnetised bar embedded in the dense star-forming centre of M82 with an ordered field morphology. The observed emission is depolarised due to the turbulent field in the star-forming regions surrounding it. This morphology might also be responsible for the two neighbouring minima in the depolarisation ratio in Fig. \ref{image_DP3cm6cm}. Since the bar is assumed to point towards the observer on the western side, the path length through the turbulent depolarising medium is minimal at its tip, leading to a local maximum in $DP$.

The timescale on which the bar has formed is important for the generation and propagation of its magnetic field and the origin of the ordered magnetic field in the halo of M82. The formation of the bar could have happenend before M82 interacted with nearby galaxies. \citet{1994Natur.372..530Y} showed that a stream of \ion{H}{i} gas connects all three galaxies in the group and suggested that M82 was a small disk galaxy before the interaction, which could also have hosted a bar. On the other hand,  \citet{1988A&A...203..259N}, for example, showed that tidal interaction induces non-circular motions, which can lead to the formation of a bar after one to two rotational timescales. This would suggest that the bar in M82 is not older than $10^8$\,Myrs, which is the timescale for the first, but not last interaction \citep{1994Natur.372..530Y}. Bars are known to originate also from tidal encounters and would only need one or two full rotations to build up a regular field via shearing. Using a value of 200\,km\,s$^{-1}$ for the speed at the tip of the bar \citep{2009RMxAC..37...44M} and a length of 1\,kpc \citep{1991ApJ...369..135T} would lead to a rotation period of $1.5\cdot 10^7$yrs, which implies that M82 would have rotated only a few times around its axis since it last interacted with M81.

One of the questions which arises when talking about the rate of star formation of M82 of 13\,M$_\odot$yr$^{-1}$ is how to fuel such a starburst consistently with new material. This is related to the question of how enough gas loses angular momentum to allow it to collapse into molecular clouds in the centre of the galaxy and create the nuclear starburst. Normally, on larger scales, magnetic fields in galaxies are not of a magnitude where they can actively influence the gas motions. The existence of a magnetised bar and a circumnuclear molecular ring in M82 enables magnetic fields to play an important role in transporting material into the star-forming centre via magnetic braking. This leads to mass inflow rates of up to 7.1\,M$_\odot$yr$^{-1}$ under optimal conditions (Sect. \ref{text_magnetic_braking}). This mass inflow rate is much greater than the molecular gas mass inflow rates derived from the CO line (3.5\,M$_\odot$yr$^{-1}$)\citep{2016ApJ...830...72C} showing that magnetic forces might play an important role in the whole star-formation process of this galaxy. Such a structure is not present in the majority of starburst galaxies and therefore cannot be the explanation for the constant fuelling of all starbursts in the Universe. 

How the large-scale magnetic field in the halo of M82 formed is difficult to determine. Strong outflows caused by the intense star formation in M82 might have expelled the magnetic field in the bar into the halo. Another explanation could be that the magnetic field is the fossil of an originally X-shaped field of the former spiral morphology of M82, which was driven further into the halo. Even though the regular field morphology in some parts looks more poloidal, following the main outflow direction, the shearing forces from the interaction with M81 and outflow might have modified the X-shape, to the extent that it can no longer be recognised. 

The problems for creating a large-scale poloidal field from dynamo action alone, namely the much smaller differential rotation in the direction of the minor axis of the galaxy, in contrast to the strong Coriolis force in the disks of galaxies, speaks for such a fossil scenario. But even for normal spiral galaxies, the generation mechanisms responsible for their strong halo fields are not well known.  

On the other hand, helical fields in the halo of M82 (Sect. \ref{text_depol}) could be the result of the magnetic field from the bar being dragged together with the gas into the halo of this galaxy. The direction of the intrinsic magnetic field (being parallel to the disk of M82 in the northern halo, but perpendicular in the southern halo) also supports the presence of helical magnetic fields. The rigidly rotating magnetic field configuration in the bar is transported into the halo where it is the subject of differential rotation, as a function of radius due to the Coriolis force as well as a function of height due to changing gas velocities, which twists it into a helix complicating the whole scenario even more. This scenario would cause strong shearing forces in the centre of the outflow cone resulting in strong  magnetic fields and therefore high RM-values which we also observed in this region (Fig.\ref{image_RM_3cm_6cm}). A helical field morphology in the central starburst region of a galaxy was also observed for NGC\,253 \citep{2011A&A...535A..79H}. Even though the evolution and mass of this galaxy differs from that of M82, and the star-formation rate in the central starburst region is a factor of $\sim3$ lower \citep{2005ApJ...629..767O}, the central star-forming nucleus shows a lot of similarities. Synthetic models from \citet{2011A&A...535A..79H} let us also conclude on the rotation direction of the magnetic field in the helix. Comparing our RMs of regions 6 ($-170$\,rad\,m$^{-2}$) and 7 ($-340$\,rad\,m$^{-2}$) (Fig. \ref{image_RM_18cm_22cm_lg_sublr}) indicates a magnetic field rotating in clockwise direction, if looking from south to north through the minor axis of M82. Since the outflow cone has a certain opening angle and our observations at $\uplambda$18\,cm/$\uplambda$22\,cm mostly trace the magnetic field at the front side of this cone, we would more likely see the azimuthal component of the helix in the northern halo and the poloidal component in the southern one. 

The asymmetry in the polarised intensities between the eastern and western halos of the galaxy could have been caused by the star-formation history of M82. Studies by \citet{1994MNRAS.266..455M} showed that the western part of the disk has been more active in the past, so that the shock wave from this star-formation episode could now have reached the halo. Using a cosmic ray bulk speed of $v=600$\,km\,s$^{-1}$ from \citet{2013A&A...555A..23A} and a distance of the structure of 1.6\,kpc would result in a travel time of $\sim$2.5\,Myrs. This is on the same order as the estimates for the last star-bursting episode from \citet{2003ApJ...599..193F} of 5\,Myrs, who used near-IR integral field spectroscopy to develop models incorporating stellar evolution, spectral synthesis, and photoionisation. This also would suggest the previous existence of a regular field, which would need to have been formed before or shortly after the starbursting period since compression of a purely turbulent field would not lead to a regular field. The compression of field lines is only able to amplify an already existing regular field. Such a compression would also be able to cause the increased polarised intensity towards the south of the galaxy where features are visible in the total power measurements \citep{2013A&A...555A..23A} as well as in infrared studies \citep{2010A&A...514A..14K}. Most of the neutral gas in the group is located towards the south of M82, which also means that the intragroup medium on this side of the galaxy is denser. This could indicate that either the generation of a large-scale regular field in the southern halo is hampered by the strong turbulence caused by the group interaction and/or the depolarisation of a regular field by a more turbulent medium in the foreground.

To what extent an already existing regular field is able to survive a violent starbursting episode is still a matter of discussion, but simulations by \citet{2009A&A...494...21A} indicate that this possibility should not be excluded.

We also see a difference in the amount of the recovered polarised emission between the northern and southern halos. A similar asymmetry has been found for the star-bursting dwarf irregular galaxy NGC1569 \citep{2010ApJ...712..536K} where less polarisation for the outflow cone towards the far side was recovered due to the inclination of the galaxy and the corresponding magnetic field morphology and depolarisation effects. Even though this galaxy is one order of magnitude smaller than M82 ($10^9$\,M$_\odot$) and at the end of its star-bursting period, it is one of the few comparable objects in the nearby Universe. It is hosting H$\upalpha$ bubbles and filaments reaching into its halo, which can be connected to its magnetic field morphology. Oppositely directed RMs within the same region were attributed to magnetic loops transported out within the warm ionised outflow. We see similar features in M82; the difference between regions 6 and 7 in Fig. \ref{image_RM_18cm_22cm_lg_sublr} and the gradient in region 9 (Fig. \ref{image_RM_18cm_22cm_lg_halo}). In both galaxies the overall polarised features are more patchy than in starburst galaxies with a more defined spiral structure \citep[e.g.][]{1997A&A...320..731D,1990A&A...240..237D,2006A&A...447..465C}. This supports the overall picture of an $\upalpha\upomega$-dynamo creating the regular magnetic field patterns in spiral galaxies and the destruction of such patterns by a violent starburst as well as the depolarisation of the radiation by turbulent fields on the line-of-sight.

\section{Summary}
\label{text_summary}

Archival data from the VLA at $\uplambda$3\,cm and $\uplambda$6\,cm and from the WSRT at $\uplambda$18\,cm and $\uplambda$22\,cm were analysed to perform a multi-frequency radio polarisation study of M82. Polarised intensities, rotation measures, and intrinsic magnetic field angles were calculated at all wavelengths. Due to the broadband multi-channel backend of the WSRT, the RM-Synthesis technique could be used to combine the $\uplambda$18\,cm and $\uplambda$22\,cm datasets and retrieve the aforementioned values without suffering from bandwidth depolarisation or the n$\pi$-ambiguity. To remove the instrumental contribution from the RM-cubes, we used a differential technique that was first proposed by \citet{2013A&A...559A..27G}, and reached dynamic ranges of $10^5$ regarding polarisation leakage.

The detected polarised structures at $\uplambda$3\,cm and $\uplambda$6\,cm comprise a large part of the western central region, which is highly polarised ($\sim11$\,\%). It is accompanied by RMs between $-60$\,rad\,m$^{-2}$ and $-180$\,rad\,m$^{-2}$ with a trough shape. The ordered magnetic field in this region is one of the strongest in the whole galaxy with $\sim11.4$\,$\upmu$G. The turbulent magnetic field has a strength of $\sim140$\,$\upmu$G. The most western part of this structure could also be recovered at the lowest noise levels with the WSRT data with low polarisation degrees (0.05\,\%), but very similar RMs. The intrinsic magnetic field orientation of this structure is disk parallel. We propose a magnetic bar as a possible explanation, which points towards the observer on the western side and away from the observer on the eastern one. This would explain the coherent magnetic field in such a violent starburst region. In addition the non-detection of polarised emission at all wavelengths on the eastern side can be explained by depolarisation due to the longer line of sight through the turbulent magnetic field in the starburst regions in front of the bar and/or a one-sided bar, which was reported earlier from \ion{H}{i}-absorption measurements.

We claim that magnetic forces are at least partially involved in moving molecular ionised gas towards the star-bursting nucleus and therefore are responsible for the high star-formation rate in M82. We calculated the possible mass-inflow rate due to magnetic braking to 7.1\,M$_\odot$yr$^{-1}$ under optimal conditions, which is even higher than the observed mass-inflow rate for the molecular gas of 3.5\,M$_\odot$yr$^{-1}$.

By using our results from the polarised spectra of this structure and due to the fact that the ionised medium is closely coupled to the magnetic field we were able to examine its magnetic field morphology. The dominant depolarisation process is Faraday Depolarisation caused by the turbulent field in the starburst region surrounding the bar and/or the turbulent field in the bar itself. The polarised emission at $\uplambda$18\,cm/$\uplambda$22\,cm is stronger than expected from a simple volume filling model. Probably, only a fraction of the available volume is filled with dense ionised gas, strengthening our claims in previous publications.

A small patch in the centre of the galaxy shows polarised emission at $\uplambda$3\,cm and $\uplambda$6\,cm. Here the intrinsic magnetic field points away from the disk, and it can be traced out to 0.6\,kpc into the halo of M82. The depolarisation ratios between those two wavelengths are inverted in the central region with values of up to 1.4. In addition larger patches of polarised emission were detected in the $\uplambda$18\,cm/$\uplambda$22\,cm RM cubes farther away from the galaxy with opposite signs of RM and magnetic field directions. This leads us to conclude that a helical magnetic field could exist in the central starbursting region of M82, perhaps extending out into the outflow cones and forming large-scale magnetic loops.

The asymmetry between the strength of the northern and southern halo fields is either the consequence of the denser environment towards the centre of the galaxy group of which M82 is a member in the south and/or the star-formation history of the galaxy.

We proposed two possible scenarios for the origin of these extended halo fields. These fields could be the remnant of the X-shaped field of M82 when it was still a spiral galaxy, before its first interaction with other nearby galaxies. Alternatively, the halo field could have originated in the bar of M82, which was then expelled and twisted by strong outflows driven by the nuclear starburst. In both cases it is most likely that the overpressure of the starburst has transported the field outwards, so that it has not been generated within the halo itself.

The detection of diffuse polarised emission at long wavelengths in the halo of a starburst galaxy and even inside the violent and most likely turbulent medium of the central region of a starburst galaxy itself is an example of how large-scale regular magnetic fields in galaxies can survive the strong starbursts in the early Universe. With the example of M82 we demonstrate a mechanism of transport of ordered magnetic field efficiently out into intergalactic space via a galactic wind. This is crucial for evolutionary models of magnetic fields in the early Universe where starbursts were more common and may be one of the main contributors to the enrichment of intergalactic space with gas.

\begin{acknowledgements}
We thank M. Westmoquette and the NOAO/AURA/NSF for providing us with the H$\upalpha$ image of the WIYN observatory (Copyright WIYN Consortium, Inc., all rights reserved) and D. Schnitzeler as well as the anonymous referee for their comments, which helped improve the manuscript. The work at the AIRUB is supported by the DFG through FOR 1048. The work at the MPIfR is supported by the DFG through FOR 1254. The Westerbork Synthesis Radio Telescope is operated by ASTRON (Netherlands Foundation for Research in Astronomy) with support from the Netherlands Foundation for Scientific Research (NWO). The National Radio Astronomy Observatory is a facility of the National Science Foundation operated under cooperative agreement by Associated Universities, Inc.
\end{acknowledgements}

\bibliography{bibtex}{}
\bibliographystyle{aa}

\listofobjects

\end{document}